\documentclass[reprint,amsmath,amssymb,aps,prx,10pt,twocolumn,groupedaddress, floatfix, superscriptaddress, longbibliography]{revtex4-1}

\usepackage{hyperref}       
\usepackage{url}            
\usepackage{booktabs}       
\usepackage{amsfonts, amsthm}       
\usepackage{nicefrac}       
\usepackage{microtype}      
\usepackage{graphicx}
\usepackage{natbib}
\usepackage{amsmath, amssymb, amsthm}
\usepackage{bm,bbm}
\usepackage{xcolor}
\usepackage{multirow}
\usepackage[export]{adjustbox}
\usepackage{color, colortbl}
\usepackage{slashed}
\usepackage{todonotes}

\let\bs\boldsymbol
\newcommand{\n}{\mathbf{n}}
\newcommand{\bpm}{\begin{pmatrix}}
\newcommand{\epm}{\end{pmatrix}}

\begin{document}

\title{Measuring the similarity of graphs with a Gaussian Boson Sampler}

\author{Maria Schuld} 
\email{maria@xanadu.ai}
\affiliation{Xanadu, Toronto, Canada}
\affiliation{University of KwaZulu-Natal, South Africa}
\author{Kamil Br\'adler}
\email{kamil@xanadu.ai}
\affiliation{Xanadu, Toronto, Canada}
\author{Robert Israel}
\affiliation{Xanadu, Toronto, Canada}
\author{Daiqin Su}
\affiliation{Xanadu, Toronto, Canada}
\author{Brajesh Gupt}
\affiliation{Xanadu, Toronto, Canada}

\date{\today}

\begin{abstract}
Gaussian Boson Samplers (GBS) have initially been proposed as a near-term demonstration of classically intractable quantum computation. We show here that they have a potential practical application: Samples from these devices can be used to construct a feature vector that embeds a graph in Euclidean space, where similarity measures between graphs - so called `graph kernels' - can be naturally defined. This is crucial for machine learning with graph-structured data, and we show that the GBS-induced kernel performs remarkably well in classification benchmark tasks. We provide a theoretical motivation for this success, linking the extracted features to the number of $r$-matchings in subgraphs. Our results contribute to a new way of thinking about kernels as a quantum hardware-efficient feature mapping, and lead to a promising application for near-term quantum computing. 
\end{abstract}

\maketitle

\section{Introduction}

Measuring the similarity of two graphs for practical applications is notoriously difficult. Firstly, there are many different notions of similarity, and practical tasks crucially depend on what property of the graph is exploited in the comparison. Secondly, even the task of determining whether two graphs are exactly the same can be computationally extremely costly. This is due to the fact that a representation of a graph is not unique: Different ways of enumerating its nodes and edges can give rise to the same structure. The complexity of deciding whether two graphs are isomorphic is unknown; neither a polynomial-time algorithm nor NP-completeness proof has been discovered yet \cite{kobler2012graph}. Existing algorithms for graph isomorphism \cite{mckay1981practical} and graph similarity \cite{ghosh2018journey} are efficient in practice, but are still costly for large graphs and may require exponential time for some problem instances.

In this paper we suggest the use of quantum hardware to map a graph $G$ to a feature vector which represents $G$ in Euclidean space. Standard distance measures, such as taking the inner product of two feature vectors, then result in a distance measure between graphs mitigated by the feature embedding. The quantum device we investigate is a Gaussian Boson Sampling (GBS) setup \cite{hamilton2017gaussian,lund2014boson, kruse2018detailed}. GBS is a generalization of Boson Sampling \cite{tillmann2013experimental,broome2013photonic}, which has originally been proposed as a classically intractable problem to demonstrate the power of near-term quantum hardware \cite{aaronson2011computational}.
An optical GBS device prepares a quantum state of $M$ optical modes and counts the photons in each mode. Some of the authors have previously shown how a graph can be encoded into the quantum state of light \cite{bradler2018gaussian}, so that the photon measurement statistics give rise to a complete set of graph isomorphism invariants \cite{bradler2018graph}. 

Here we extend this result and study the graph similarity measure derived from a GBS device for a practical application, namely for classification for machine learning. Graph-structured data plays an increasingly important role in this field, for example to predict properties of a social media network given a dataset of networks for which the properties are known.  In machine learning, a similarity measure between data is called a \textit{kernel}, and lots of methods for pattern recognition -- such as support vector machines and Gaussian processes -- are built around this concept. Mapping graphs to feature vectors or graph embeddings \cite{zhang2018network, goyal2018graph, grover2016node2vec} is a well known strategy, and graph kernels from explicit feature vectors \cite{kriege2014explicit} have been studied in detail.

The connection between kernel methods for machine learning and quantum computing has recently been made in Refs. \cite{schuld2019quantum, havlivcek2019supervised}. Any positive-definite kernel can be formally understood as the inner product of two feature vectors that represent the data points in a Hilbert space \cite{scholkopf2001learning}. 
Hence, the Hilbert space of a quantum system can be interpreted as a feature space, in which a subroutine can compute inner products ``coherently''. By using measurement samples from the quantum hardware to construct low-dimensional feature vectors that can be stored and further processed on a classical computer, we follow a different, even more minimalistic route to define a ``quantum feature map'', and ultimately a quantum kernel. The advantage in using quantum hardware this way is that device performs a combinatorial computation that is very resource-intense -- possibly even intractable -- for classical computers. In fact, we show that the GBS feature map is related to a class of classical graph kernels which count subgraphs \cite{shervashidze2009efficient}, but instead of only considering subgraphs of constant size, the sampling statistics reveal information on all possible subgraphs, as well as subgraphs constructed from copying nodes and their edges. The resulting features contain information about the number of $r$-matchings of the original graph. Numerical experiments reveal that graph kernels from a GBS-induced feature map can outperform classical graph kernels in classification task for standard benchmark datasets, results that can be further improved by using displaced light modes.

\begin{figure*}[t]
\centering
\includegraphics[width = 0.75\textwidth]{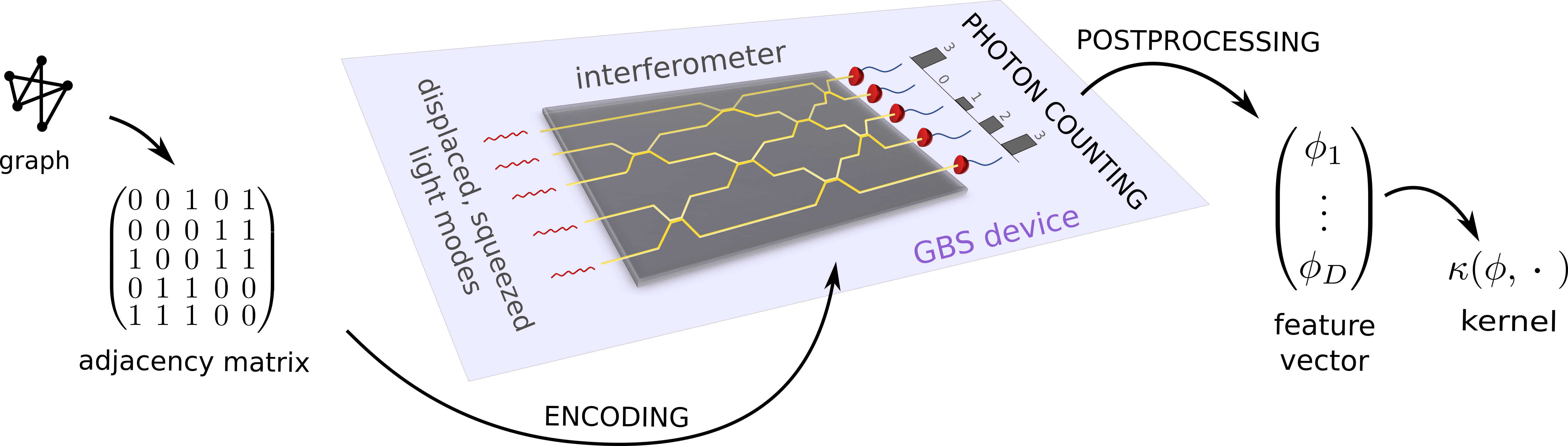}
\caption{Idea of the quantum hardware-induced feature map. The adjacency matrix of a graph gets encoded into the Gaussian state of the light modes by tuning the squeezing and interferometer parameters. The features are defined as the probability of detecting certain classes of photon counting events. To extract the probabilities from the device, a number of samples of photon counting events is generated and the relative frequencies of the different classes determined. 
}
\label{Fig:gbs}
\end{figure*}

\section{Turning GBS samples into features}
\label{Sec2}

An optical Gaussian Boson Sampler is a device where a special quantum state (a so-called \textit{Gaussian state}) is prepared by the \textit{optical squeezing} of $M$ \textit{displaced} light modes, followed by an \textit{interferometer} of beamsplitters.
Such a Gaussian state is fully described by a covariance matrix $\sigma \in \mathbb{R}^{2M\times2M}$ as well as a displacement vector $\mathbf{d} \in \mathbb{R}^{2M}$ \cite{weedbrook2012gaussian}. Photon number resolving detectors count the photons in each mode.

In this section we describe the mathematical details of the quantum hardware-induced feature map (see also Figure~\ref{Fig:gbs}), summarizing what has been described in Br{\'a}dler et al. \cite{bradler2018graph}, and adding the effect of displacement as well as a further step of turning samples to feature vectors through what we will call ``meta-orbits''. The scheme works for \textit{simple graphs}, i.e. undirected graphs without self-loops or multiple edges. While edge weights can be treated on the same footing as unweighted edges, we leave the inclusion of categorical edge labels or node labels for future studies. Mindful of readers from fields other than quantum optics we will only highlight some important aspects of Gaussian Boson Sampling and refer to Refs. \cite{hamilton2017gaussian, kruse2018detailed, bradler2018gaussian}  for more detail.

\subsection{Encoding graphs into the GBS device}
\label{Sec:enc}

As outlined in \cite{bradler2018gaussian}, a quantum state prepared by a GBS device can encode a graph $G = (V, E)$ with an adjacency matrix $A$ of entries $A_{ij}$ that are one if the edge $(i,j)$ exists in $G$ and zero else.
The entries of $A$ can also represent continuous ``edge weights'' that denote the strength of a connection. In the latter case we will speak of a ``weighted adjacency matrix''.

In order to associate $A$ with the symmetric, positive definite $2M$-dimensional covariance matrix of a Gaussian state of $M$ modes, we have to construct a ``doubled adjacency matrix''
\begin{equation}
\tilde{A} = c\begin{pmatrix} A & 0 \\ 0 & A\end{pmatrix} = c(A \oplus A),
\label{Eq:doubleA}
\end{equation}
where the rescaling constant $c$ is chosen so that $0 < c < 1/s_{\mathrm{max}}$, and $s_{\mathrm{max}}$ is the maximum singular value of $A$ \cite{bradler2018graph, bradler2018gaussian}.\footnote{As long as it fulfills the above inequality, $c$ can be treated as a hyperparameter of the feature map, which may also be influenced by hardware constraints since it relates ultimately to the amount of squeezing required.} For simplicity we will always rescale all adjacency matrices with a factor $1/(s^{\{G\}}_{\mathrm{max}} + 10^{-8})$ where $s^{\{G\}}_{\mathrm{max}}$ is the largest singular value among all graphs in the data set under consideration.
As a result we will assume that $c=1$ and $\tilde{A} = A \oplus A$ can be encoded into a GBS device. We call this the ``doubled encoding strategy''.

The matrix $\tilde{A}$ can now be associated with a quantum state's covariance matrix $\sigma$ by setting the squeezing as well as the beamsplitter angles of the interferometer so that
\begin{equation}
	\sigma  =  Q -  \mathbbm{1}/2, \text{ with } Q = (\mathbbm{1} - X\tilde{A})^{-1} , X = \begin{pmatrix} 0 & \mathbbm{1} \\ \mathbbm{1} & 0 \end{pmatrix}.
	\label{Eq:Q}
\end{equation}

\subsection{Sampling photon counting events}

\begin{table}[t]
\centering
\includegraphics[scale=1]{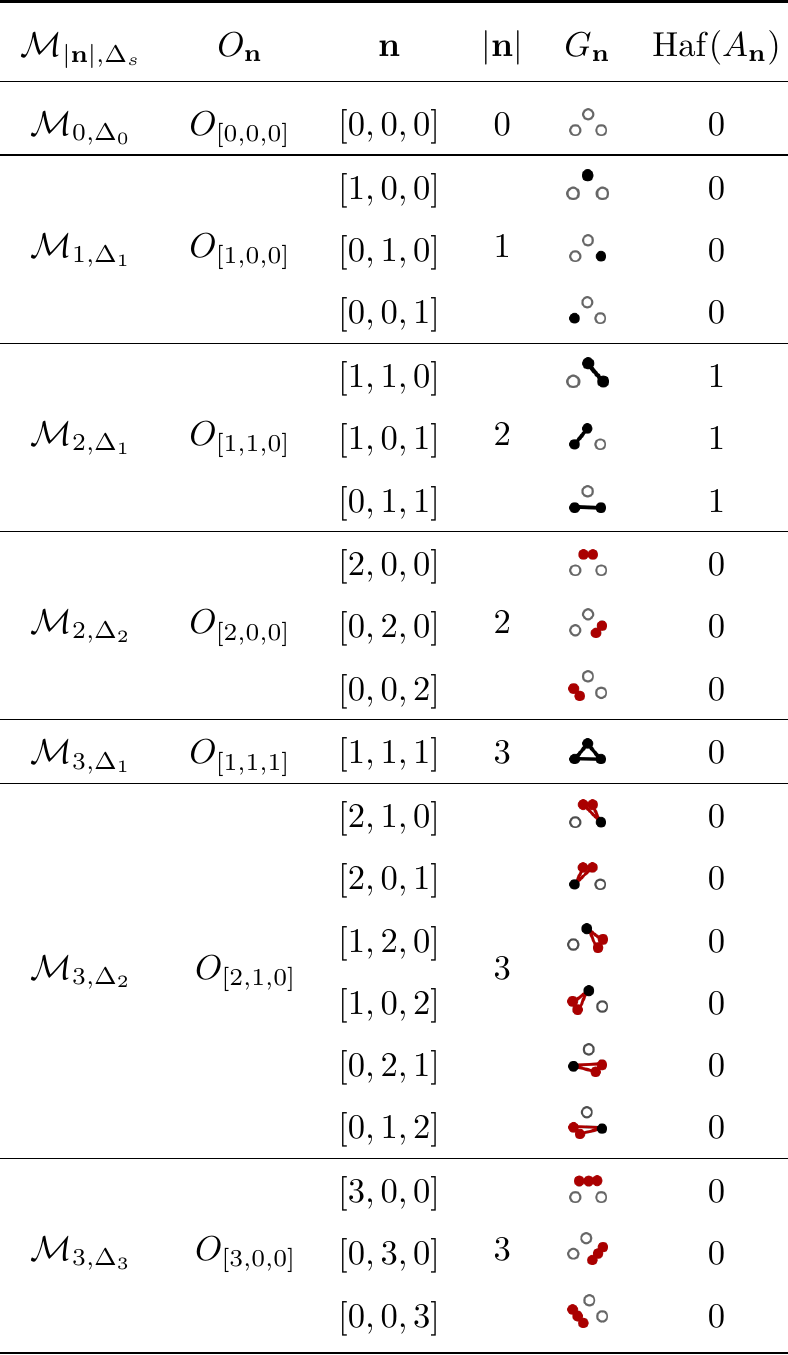}
\vspace{0.3cm}
\caption{Meta-orbits $\mathcal{M}_{|\n|,\Delta_{s}}$, orbits $O_{\n}$, photon events $\n$, total photon number $|\n|$, extended induced subgraph $G_{\n}$ (indicated by red/black nodes and edges) and Hafnian $\mathrm{haf}(A_{\n})$ of a fully connected simple graph of three nodes up to $|\n|_{\mathrm{max}} = 3$. The difference between orbits and meta-orbits only becomes apparent for higher photon evens (i.e., $[2,2,0,0]$ and $[2,1,1,0]$ are in different orbits but the same meta-orbit). Note that the red nodes are not mutually connected. }
\label{Tbl:3nodes}
\end{table}

After embedding $A$ via $\tilde{A}$ into the quantum state of the GBS, each measurement of the photon number resolving detectors returns a photon event $\mathbf{n} = [n_1, \dots, n_M]$, with $n_i\in\mathbb{N}$ indicating the number of photons measured in the $i$-th mode.
Assuming for now that the displacement $\mathbf{d}$ is zero, the probability of measuring a given photon counting event is
\begin{equation}
	p(\n ) = \frac{1}{\sqrt{\mathrm{det}(Q)} \; \n! } \; \mathrm{haf}^2 (A_{ \n }) ,
	\label{Eq:distribution}
\end{equation}
where $\n ! = n_1! n_2! \cdots n_M!$.

Let us go through this nontrivial equation bit by bit. The Hafnian $\mathrm{haf}()$ is a matrix operation similar to the determinant or permanent. For a general symmetric matrix $C \in \mathbb{R}^{N} \times \mathbb{R}^{N}$ it reads
\begin{equation} \mathrm{haf}(C) = \sum_{\pi \in P_{N}^{\{2\}}} \prod_{\scriptscriptstyle (u, v) \in \pi} C_{u, v}.
\label{Eq:partition1}
\end{equation}
Here, $P_{N}^{\{2\}}$ is the set of all $N!/((N/2)!2^{N/2})$ ways to partition the index set $\{1, 2, \dots, N \}$ into $N/2$ unordered pairs of size $2$, such that each index only appears in one pair. The Hafnian is zero for odd $N$.
As an example, for the index set $\{1,2,3,4\}$ we have $P_{4}^{\{2\}} =  \{ (1,2), (3,4)\}, \{ (1,3), (2,4)\}, \{ (1,4), (2,3)\} $.

\begin{figure*}
\includegraphics[width = \textwidth]{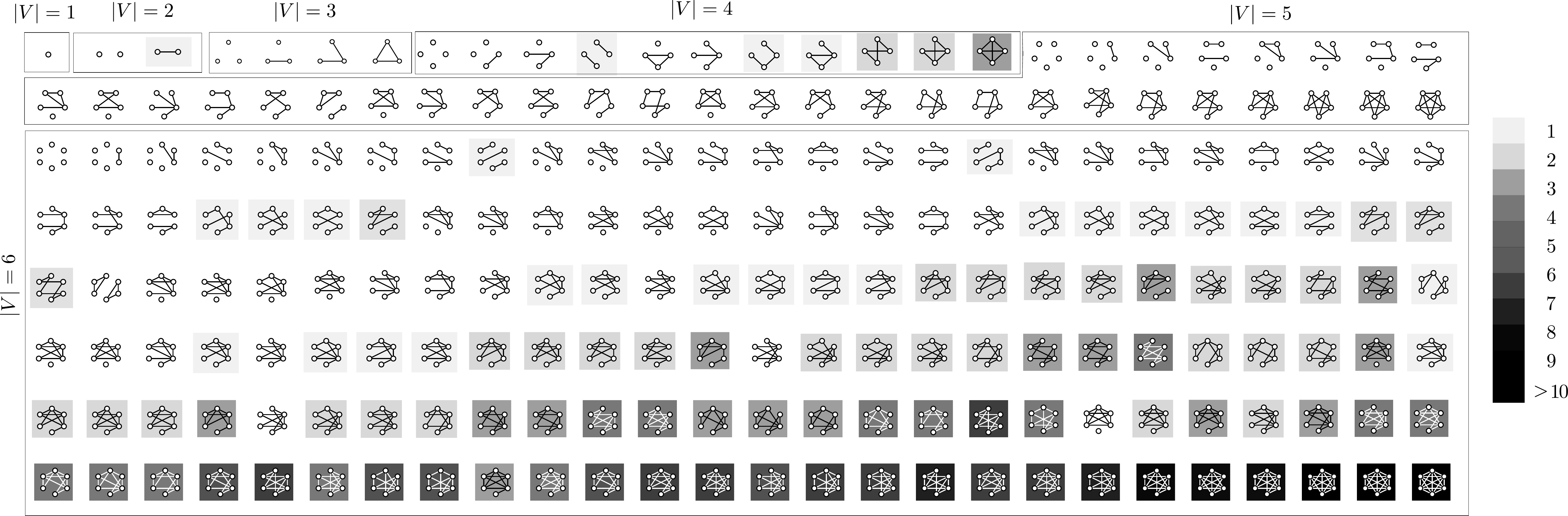}
\caption{All non-isomorphic graphs up to size $|V|=6$ and the number of perfect matchings they contain (grey shading scale). }
\label{Fig:all}
\end{figure*}

If $C$ is interpreted as an adjacency matrix containing the edges of a graph, the set $P_{N}^{\{2\}}$ contains edge-sets of all possible perfect matchings on $G$.
A perfect matching is a subset of edges such that every node is covered by exactly one of the edges. 
The Hafnian therefore sums the products of the edge weights in all perfect matchings.
If all edge weights are constant, it simply counts the number of perfect matchings in $G$ (see also Figure \ref{Fig:all}).
Note that in Eq.~(\ref{Eq:distribution}) we used the fact that for real and symmetric $A$, $\text{haf}(\tilde{A}) = \text{haf}(A \oplus A) = \mathrm{haf^2}(A)$.
In other words, the doubled encoding strategy leads to a square factor which will play a profound role in the quantum feature map we are aiming to construct.

Eq.~(\ref{Eq:distribution}) does not depend on the Hafnian of the adjacency matrix $A$, but on a matrix $A_{ \n }$. $A_{ \n }$ contains $n_j$ duplicates of the $j$th row and column in $A$. If $n_j=0$, the $j$th row/column in $A$ does not appear in $A_{ \n }$. Effectively, this constructs a new graph $G_{\n}$ from $A$ according to the following rules (see also Table~\ref{Tbl:3nodes}): 
\begin{enumerate}
\item If all $n_j, j=1,\dots,M$ are one (i.e., each detector counted exactly one photon), $A_{\n} = A$. 
\item If some $n_j$ are zero and others one (i.e., these detectors report no photons), $A_{\n}$ describes an \textit{induced subgraph} $G_{\n}$ of $G$, in which nodes that correspond to detectors with zero count were deleted together with any edge that connected them to other nodes. 
\item If some $n_j$ are larger than one (i.e., these detectors count more than one photon), $A_{\bs{N}}$ describes what we call an \textit{extended induced subgraph} in which the corresponding nodes and all their connections are duplicated $n_j$ times.
\end{enumerate}
In short, the probability of a photon event to be measured by the GBS device is proportional to the square of the (weighted) number of perfect matchings in a -possibly extended - induced subgraph of the graph encoded into the interferometer. 

Computing the Hafnian of a general matrix is in complexity class \#P, and formally reduces to the task of computing permanents \cite{valiant1979complexity}. If no entry in the matrix is negative, efficient approximation heuristics are known, although their success is only guaranteed under specific circumstances \cite{barvinok2016approximating, rudelson2016hafnians}. 

\subsection{The effect of displacement}

The Gaussian Boson Sampling setup underlying Eq.~(\ref{Eq:distribution}) consists of squeezing and interferometers. But a Gaussian quantum state can also be manipulated by a third operation: displacement. Displacement changes the mean of the $M$-mode Gaussian state while leaving the covariance matrix (and therefore the encoding strategy) as before. A non-zero mean changes Eq.~(\ref{Eq:distribution}) in an interesting, but non-trivial manner.

Without going into the details \cite{kruse2018detailed}, if considering nonzero displacement, instead of summing over $P_{N}^{\{2\}}$ in Eq.~(4), we have to sum over $P_{N}^{\{1,2\}}$, or the set of partitions of the index set $\{1,\dots,N\}$ into subsets of size \textit{up to} $2$.
For the index set $\{1,2,3,4\}$, we had 
$$P_{4}^{\{2\}} =  \{ (1,2), (3,4)\}, \{ (1,3), (2,4)\}, \{ (1,4), (2,3)\}, $$ 
which now becomes
{\small
\begin{align*}
P_{4}^{\{1,2\}} = \{ &  \{ (1,2), (3), (4)\}, \{ (1,3), (2), (4)\}, \{ (1,4), (2), (3)\}, \\
&\{ (2,3), (1), (4)\}, \{ (2,4), (1), (3)\}, \{ (3,4), (1), (2)\}, \\
& \{ (1,2), (3,4)\}, \{ (1,3), (2,4)\}, \{ (1,4), (2,3)\},\\
& \{ (1), (2), (3), (4)\}
\end{align*}
}
Instead of the Hafnian in Eq.~(\ref{Eq:distribution}), we therefore get a mixture of Hafnians of $A_{\n}$'s submatrices (stemming from the pairs) and other factors (stemming from the size-1 sets).

Assume that displacement is applied to both the $\hat{x}$ and $\hat{p}$ quadratures of each mode, described by a $2M$-dimensional displacement vector $\mathbf{d} = (d_1,\dots,d_M, d_{1}^*,\dots,d_{M}^*)^T$. The effect on Eq.~(3) is as follows.
Let $Q$ the $2M \times 2M$ matrix from Eq.~(\ref{Eq:Q}), and $\mathbf{b}=\textbf{d}^{\dagger} Q^{-1}$. We get\footnote{To derive Eq.~(\ref{Eq:displ}) from the analysis in \cite{kruse2018detailed}, one uses the fact that for $\tilde{A}$ being a direct sum $A \oplus A$, the index set $i_1,\dots,i_{2n} \in \mathcal{I}_{2M}$ can be divided into two index sets: $j_1,\dots,j_s$ which contains all $s$ indices from the `first subspace' (i.e., the first $M$ dimensions) of $\tilde{A}$, and $k_1,\dots, k_{s'}$ containing the $s'$ indices from the `second subspace', with $s+s' = 2n$. The fact that $\mathrm{haf}(A \oplus B) = \mathrm{haf}(A)\mathrm{haf}(B)$, allows us to express the Hafnian of reduced versions of $\tilde{A}_{\n} $ as a product of reduced versions of matrix $\tilde{A}_{\n} $,
$$ \mathrm{haf}(\tilde{A}_{\n - \{i_1,\dots,i_{2n}\}}) = \mathrm{haf}(A_{\n - \{j_1,\dots,j_{s}\}}) \mathrm{haf}(A_{\n - \{k_1,\dots,k_{s'}\}}). $$}
\begin{equation}
	p(\n ) = \alpha
	 \left[ \sum\limits_{n = 0}^{M} \sum\limits_{\{i_1..i_{n}\} \subseteq \mathcal{I}_{M} }
	b_{i_1} \hdots b_{i_{n}} \mathrm{haf}(A_{\n - \{i_1..i_{n}\}})\right]^2
	\label{Eq:displ}
\end{equation}
with 
$$  \alpha = \frac{e^{-\frac{1}{2} \mathbf{d}^{\dagger} Q^{-1} \mathbf{d}}}{\sqrt{\mathrm{det}(Q)} \; \n! },$$
where $\mathcal{I}_{2M}$ is the index set $\{1,\dots 2M\}$. In this notation we assume $\{i_1,\dots, i_0\} = \{\}$ and $b_{i_1} \dots b_{i_0}=1$.
The ``reduced'' Hafnians of the form $ A_{\n - \{i,j\}},  A_{\n - \{i,j,k,l\}}\dots$ are constructed by ``deleting'' rows and columns $\{i,j\}, \{i,j,k,l\},...$ in $\tilde{A}_{\n}$. The expression in the brackets of Eq.~(\ref{Eq:displ}) is also known as a ``loop Hafnian'' of a matrix $\tilde{A}_{\n}$ that carries $b_1,...,b_{2M}$ on its diagonal \cite{quesada2019franck}.

One can see that displacement explores substructures of extended subgraphs, adding another layer of ``resolution'' to the photon number distribution. An important effect of displacement is that $p(\n)$ for odd total photon numbers $|\n|$ is not necessarily zero any more, since the sum in Eq.~(\ref{Eq:displ}) contains Hafnians of even-sized subgraphs.

\subsection{Turning samples into features}

The basic idea of how to turn samples of photon counting events into feature vectors is to associate the probability of a certain measurement result with a feature. To estimate the probability of measurement outcomes, one simply divides the number of times a result has been measured by the total number of measurements.
However, if we simply used the probabilities of photon events $p(\n)$ as features, we would face a very fast -- more precisely, a doubly factorial -- explosion of the number of features with the total number of photons, while almost all events become vanishingly unlikely for realistic amounts of squeezing.
In practice we will truncate the total number of photons at a fixed value $k$ and discard all measurement results with $|\n| > k$ in the construction of the feature vector, but even then the sampling task quickly becomes unfeasible. 

We therefore define the probability of certain \textit{types} of photon events as features, thereby ``coarse-graining'' the probability distribution.
As a compromise between experimental feasibility and expressive power, we consider two different coarse-graining strategies here. The first one  follows Br{\'a}dler et al.'s \cite{bradler2018graph} suggestion  to coarse-grain the distribution of photon counting events by summarizing them to sets called \textit{orbits} (see Table~\ref{Tbl:3nodes}).
An orbit $O_{\n} = \{\text{perm} (\mathbf{n}) \}$ contains permutations of the detection event $\mathbf{n}$. For example, $[2,1,1,0]$ is in the same orbit as $[0,1,2,1]$, but not $[2,2,0,0]$. The photon counting event $\n$ in the index is therefore an arbitrary ``representative'' of the photon counting events in an orbit.
The probability of detecting a photon counting event of orbit $O_{\n}$ is given by the sum of the individual probabilities,
\begin{equation}
	p(O_{\n} ) :=  \sum\limits_{\n \in O_{\n}} p(\n).
	\label{Eq:orbit}
\end{equation}
The number of orbits $O_{\n}$ containing events of up to $k$ photons in total is equal to the number of ways that the integers of $1,...,k$ can be partitioned into a sum of at most $M$ terms. In practice we usually have $k\ll M$, in which case there are $2,4,7,12,19,30,45,67$ orbits for $k=1,\dots,8$, respectively\footnote{See also A000070 in the Online Encyclopedia of Integer Sequences, \url{https://oeis.org/A000070}.}. 
In a real GBS setup, the energy is finite and high photon counts therefore become very unlikely. \footnote{The energy of a Gaussian quantum state, and hence the average photon number, is determined by the squeezing and displacement operations.}

The second post-processing strategy builds on top of the first, and summarizes orbits to `meta-orbits' $\mathcal{M}_{|\n|,\Delta_{s}}$,
where
$$
\Delta_s=\big\{\n:\sum_{i}n_i=|\n|,(\forall i)(n_i\leq s) ,(\forall\n\exists n_i\in\n)(n_i=s)\big\}.
$$
In words, a meta-orbit contains all orbits of $|\n|$ photons, which have at least one detector counting $s$ photons, but no detector counts more than $s$ photons (see also Table \ref{Tbl:3nodes}).
The probability of detecting an event from a meta-orbit is given by
\begin{equation}
      p(\mathcal{M}_{|\n|,\Delta_{s}}) := \sum_{\n \in \Delta_s} p(O_{\n}).
      \label{Eq:DeltaCoarseGr}
\end{equation}
From here on, when using meta-orbit features, we refer to the GBS as ``GBS$^+$''.

It is interesting to estimate how many samples are needed to estimate a feature vector. In Ref \cite{shervashidze2009efficient} we find that we can approximate a probability distribution of $D$ possible outcomes, with 
probability at most $\delta$ that the sum of absolute values of the errors in the empirical probabilities 
of the outcomes is $\epsilon$ or more, using
\[
S = \left\lceil \frac{2(\log (2)D + \log (\frac{1}{\delta}))}{\epsilon^2}\right\rceil
\]
samples.
For orbits up to $k=8$ photons, there are $D = 67$ features. Setting $\epsilon = 0.05$ and $\delta=0.05$ and assuming a perfect GBS device, we need $39,550$ samples. Since current-day photon number resolving detectors can accumulate about $10^5$ samples of photon counting events per second \cite{vaidya2019broadband}, it takes in principle only a fraction of a second for the orbit probabilities to be estimated by the physical hardware. The number of samples does not grow with the graph size, but of course the GBS device itself grows linearly in the number of nodes. 

While hardware implementations of Gaussian Boson Samplers are rapidly advancing, in this paper we still resort to simulations. 
Sampling from photon event distributions is still a topic of active research, and to ensure that the results are not influenced by approximation errors we will use exact calculations here. This limits the scope of the experiments to graphs of the order of $25$ nodes.

\subsection{Constructing a similarity measure}

Summarizing the above, the feature map implemented by a GBS device maps a graph to a feature vector, $G \rightarrow \mathbf{f} \in \mathbb{R}^D$, where the entries $f_i, i=1,\hdots,D$ of $\mathbf{f}$ are the probabilities of detecting certain types of photon events that we called orbits and meta-orbits,
\begin{equation}
	f_{i} = p(O^i_{\n}), \text{ or } f_{i} = p(\mathcal{M}^i_{|\n|, \Delta_s}),
	\label{Eq:featmap}
\end{equation}
and the probability of the $i$'th (meta-)orbit is fully defined by Eqs.~(\ref{Eq:orbit}) and~(\ref{Eq:DeltaCoarseGr}) (while ordering in the feature vector does not matter).

Assuming that the maximum number $k$ of photons we consider is smaller or equal to the number of detectors, or $k\leq M$ for all graphs, the size $D$ of the feature vector is solely determined by $k$, which is a hyperparameter of the feature map. Another hyperparameter is the displacement that can be applied to the light modes. We will assume here that the displacement applied to all modes is a constant value $d$.

Once constructed, the feature vectors can be used for various applications. In the context of machine learning, they can be directly fed into neural network classifiers. Here we are interested in constructing a similarity measure or kernel that computes the similarity between two graphs $G$ and $G'$. A standard choice is to use the feature vectors in a `linear' and `rbf' kernel (with a hyperparameter $\delta$) 
\begin{align*}
\kappa_{\mathrm{lin}}(G , G') &=  \langle \mathbf{f}, \mathbf{f}' \rangle,\\
\kappa_{\mathrm{rbf}}(G , G') &=  \exp \left({-\frac{||\mathbf{f} - \mathbf{f}'||^2}{2\delta^2}}\right),
\end{align*}
both of which are well known to be positive semi-definite so that the results of kernel theory apply to the ``GBS kernel'' constructed here.

\section{The GBS graph features}
\label{Sec3}

In this section we will analyze the features of the first post-processing strategy in more detail; we discuss their intimate relation to the coefficients of a graph property called a ``matching polynomial'', the relation of photon event probabilities to higher-order moments of multivariate normal distributions, the connection between the GBS and graphlet sampling kernel, and we finally discuss the devastating effect of photon loss on the features.

\subsection{Single-photon features and $r$-matchings}

\begin{figure}[t]
\centering
\includegraphics[width=0.3\textwidth]{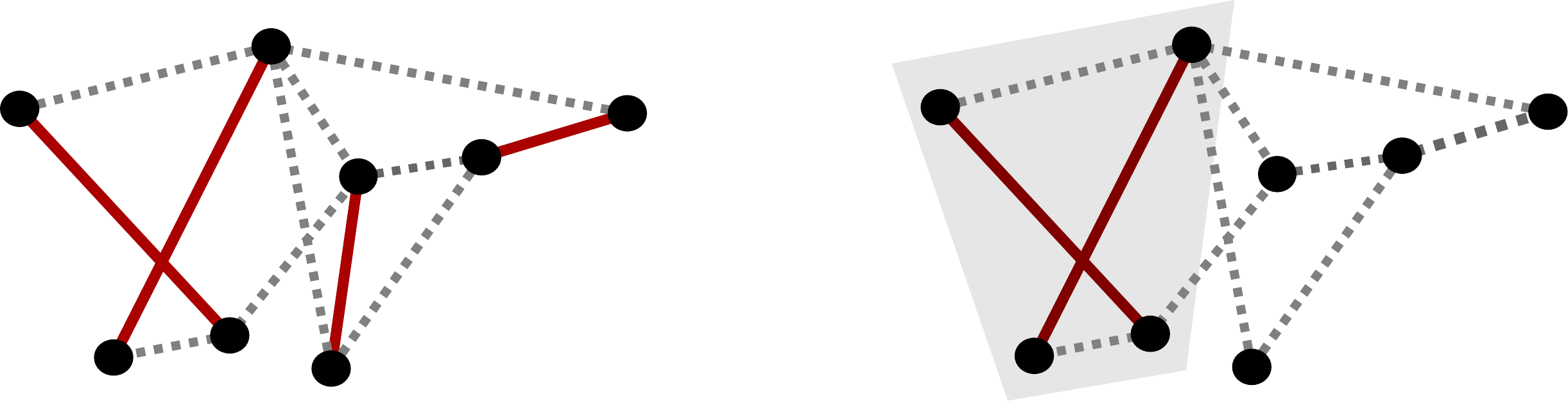}
\caption{Example of a perfect matching (left) and a $2$-matching (right). The $2$-matching is at the same time  a perfect matching of the subgraph highlighted in grey. }
\label{Fig:fig2a}
\end{figure}

It turns out that the probabilities of `single-photon' orbits (i.e., each detector counts either zero or one photon) are related to a graph property called the ``matching polynomial'' of $G$ \cite{farrell1979an,godsil1981on,heilmann1972theory},
\begin{equation} \mu(G) = \sum\limits_{r=0}^{\lceil M/2 \rceil} (-1)^r m(G,r) x^{M-2r}. 
\label{Eq:mp}
\end{equation} 
The coefficients $m(G,r)$ of the matching polynomial count the number of $r$-matchings or ``independent edge sets'' in $G$ -- sets of $r$ edges that have no vertex in common (see Figure~\ref{Fig:fig2a}).
 In the language of Hafnians, the $r$ matching can be written as $m(G,r) = \sum_{\n \in O_{[1,\dots,1,0, \dots]}} \mathrm{haf}(A_{\n})$ (where $[1,\dots,1,0, \dots]$ contains $2r$ single photon detections).
Hence, if it were not for the square of the Hafnian in Eq.~(\ref{Eq:distribution}), the probability $p(O_{\n})$ of a single-photon orbit would be proportional to a $|\n|/2$-matching $m(G,|\n|/ 2 )$ of $G$.
The square gives rise to a new object 
$$g(G,r) = \sum_{\n \in O_{[1,\dots,1,0, \dots]}} \mathrm{haf}^2(A_{\n}).$$
Replacing $m$ with $g$ in Eq.~(\ref{Eq:mp}) leads to a new type of polynomial $\gamma(G)$ which we call a \textit{GBS polynomial}. 

This definition opens up a range of interesting questions, for example whether the GBS polynomial has advantages over a standard polynomial, or how multi-photon events and displacement fits into this interpretation. We will investigate these questions in separate works.

An interesting observation for the context of machine learning occurs for the feature corresponding to orbit $O_{[1,1,0,\dots]}$ (see for example Table~\ref{Tbl:3nodes}).
Since there are only two options -- the two nodes are connected and have therefore exactly one perfect matching, or they are not and have none -- the square does not have any effect, and the probability of the orbit is proportional to the number of $1$-matchings of this graph, which is in turn equal to its number of edges.
Hence, we have that $p(O_{[1,1,0,\dots]}) \propto |E|$, and the hardware natively returns an ``edge counting'' feature.

\subsection{Higher-order moments}

The probability of measuring a given photon counting event $\n =[n_1,\dots,n_M]$ can also be interpreted from a slightly different, more physically motivated viewpoint.
The $M$ nodes of a graph can be associated with $M$ random variables drawn from a multivariate normal distribution $N(\xi, \Sigma)$, where the covariance matrix $\Sigma$ corresponds to the doubled adjacency matrix $\tilde A$, and $\xi$ is the mean vector related to displacement via $\xi = Q^{-1} \mathbf{d}^{\dagger}$.
The higher-order moments $ E[X_1^{(1)}\dots X_1^{(n_1)}\dots X_M^{(1)}\dots X_M^{(n_M)}  ]$ of this distribution are proportional to $\mathrm{haf}(A_{\n})$, which in turn is related to the probability of a photon event via Eq.~(\ref{Eq:distribution}).
This result follows from \emph{Isserlis' theorem}~\cite{isserlis1918formula}, which decomposes the higher order moments into sums of products of covariances $E[X_a X_b]$. In short, the GBS device turns a graph into a multivariate normal distribution and samples from its moments.

Using this picture, the first-order moments of the `graph-induced distribution' correspond to photon events of the form $[1,0,\dots]$ and their probability is indeed proportional to the mode means as apparent from Eq.~(\ref{Eq:displ}).
The second-order moments correspond to photon events of the form $[1,1,0,\dots]$ and their probability is proportional to the entries of the adjacency matrix -- the edge weights.
Consistent with this observation, we stated before that orbits with $2$ non-zero detectors ``measure'' the edge count of a graph.

While the doubled encoding strategy as well as the presence of multi-photon events somewhat obscure interpretations of features in terms of $r$-matchings and higher-order moments, we found in numerical experiments not reported in this paper that they can be a blessing in disguise, making very similar graphs distinguishable by smaller maximum photon numbers $k$.

\subsection{Comparison to Graphlet Sampling kernel}

Counting subgraphs in a larger graph is a concept used in various classical graph kernels. Graphlet Sampling kernels \cite{shervashidze2009efficient} bear the most striking similarity to GBS feature maps, since the features count how often graphlets of size $|V|=3,4,5,\dots$ appear in a graph $G$. In the language developed here we can express the feature $f_g$ which counts graphlet $g$ via
\begin{equation}
	 f_g \propto \sum\limits_{\n \in O_{[1,..,1,0,\dots]}}  \mathbbm{1}_{g \cong G_{\n}}, 
	\label{Eq:pm_1}
\end{equation}
using an indicator function $\mathbbm{1}_{g \cong G_{\n}}$ that is one if graphlet $g$ is isomorphic to the subgraph $G_{\n}$ and zero else, as well as the orbit represented by $[1,..,1,0,\dots]$ counting $|V|$ single photons.
In comparison, rewriting  Eq.~(\ref{Eq:featmap}) in a similar way, the GBS features are
\begin{equation}
	f_i = f_{\n_i^*} \propto \sum_{\n \in O_{\n^*}} \left( \sum\limits_{g \in \mathcal{P}^{|\n|}}  \mathbbm{1}_{g \cong G_{\n}} \right)^2  , \tag{S4}
	\label{Eq:pm_2}
\end{equation}
where $\mathcal{P}^{|\n|}$ is the set of all perfect matchings of size $|\n|$.
As a result, instead of counting graphlets, the GBS feature map sums squares of perfect matching counts in graphlets. Also, GBS feature map does not restrict the size of the graphlet probed.

\subsection{Errors due to photon loss}

\begin{figure}[t]
     \includegraphics[width=0.5\textwidth]{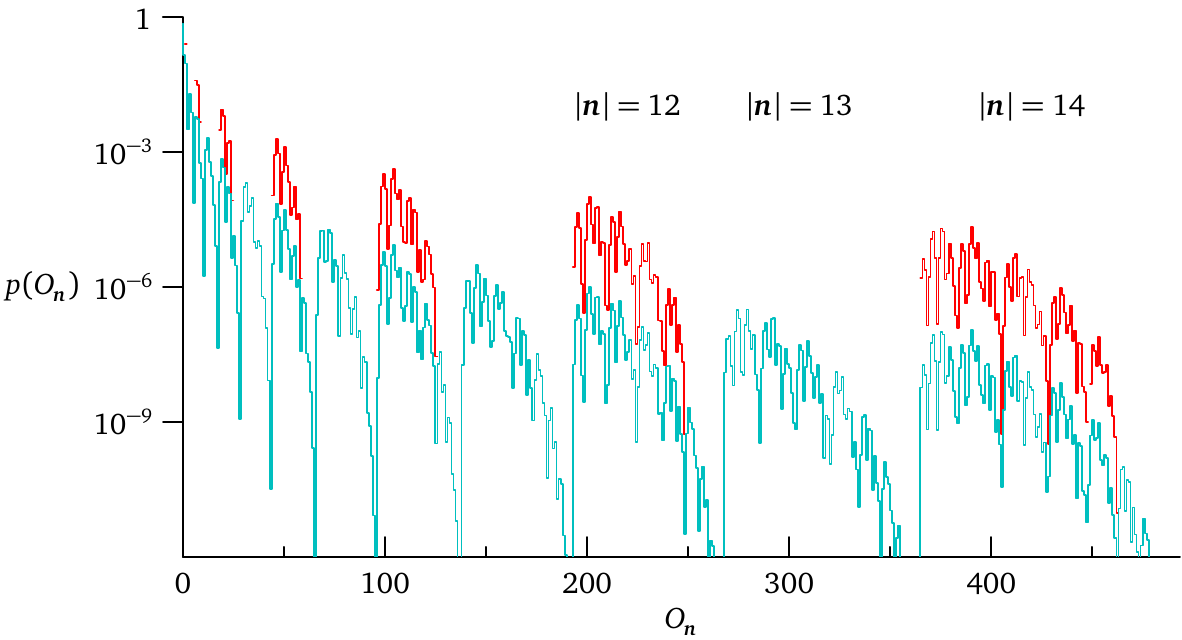}
      \caption{Coarse-grained probability $p(O_{\n})$ where $G$ is an random unweighted graph on $10$ vertices. We compare the lossless scenario (red) with a lossy case (blue) of $3$dB photon loss ($\nu=0.5$ in~\eqref{eq:lossyChannel}). The squeezing is the same in both cases and its maximal value is 6.2dB. The orbits are ordered on the $x$ axis by sorting the detectors by the size of their photon numbers.}
      \label{Fig:loss}
\end{figure}

One of the main sources of errors in a realistic GBS device is a photon loss in the linear interferometer, and we demonstrate here that loss is a serious problem for applications of a GBS for graph similarity as proposed in this paper. Methods of dealing with this kind of errors will be discussed in upcoming work. Here we show the effect of the loss on the coarse-grained probabilities with a numerical example. 

The effect of loss is described by the action of the lossy bosonic channel on a pure covariance matrix $\sigma$ resulting in
\begin{equation}\label{eq:lossyChannel}
  \sigma(\nu)=(1-\nu)\sigma+{\frac{\nu}{2}}\mathbbm{1}_M,
\end{equation}
where $\nu=1-\eta$ and $\eta$ is the overall transmissivity. 
One way of viewing this is that the matrix $\tilde{A}$ from Eq.~(\eqref{Eq:doubleA}) does not have the block-diagonal structure $c(A\oplus A)$ any more, but is of the form
\begin{equation*}\label{eq:Ceps}
  \tilde{C}=X_{2M}\big(\mathbbm{1}_{2M}-W^{-1}\mathrm{diag}\left[{\Lambda_1-1\over\Lambda_1\nu-1},\dots,{\Lambda_{2M}-1\over\Lambda_{2M}\nu-1}\right]W\big),
\end{equation*}
where $WX_{2M}\tilde{A}W^{-1}=\bs{\Lambda}$ is the eigendecomposition of $X_{2M}\tilde{A}$. 
Figure \ref{Fig:loss} shows the effect of this loss model on the probability distribution $p(O_{\n})$ over orbits for a random unweigthed graph $G$ on ten vertices.
It is apparent that loss introduces errors in the distribution, populating orbits which have a zero probability in the zero-displacement case, and distorting the remaining probabilities significantly. 
In the remainder of the paper we will consider only a lossless GBS device, but remark herewith that loss mitigation strategies are crucial for practical applications of GBS feature maps.

\section{Experiments}
\label{Sec4}

\begin{figure*}[t]
\centering
\includegraphics[width=0.75\textwidth]{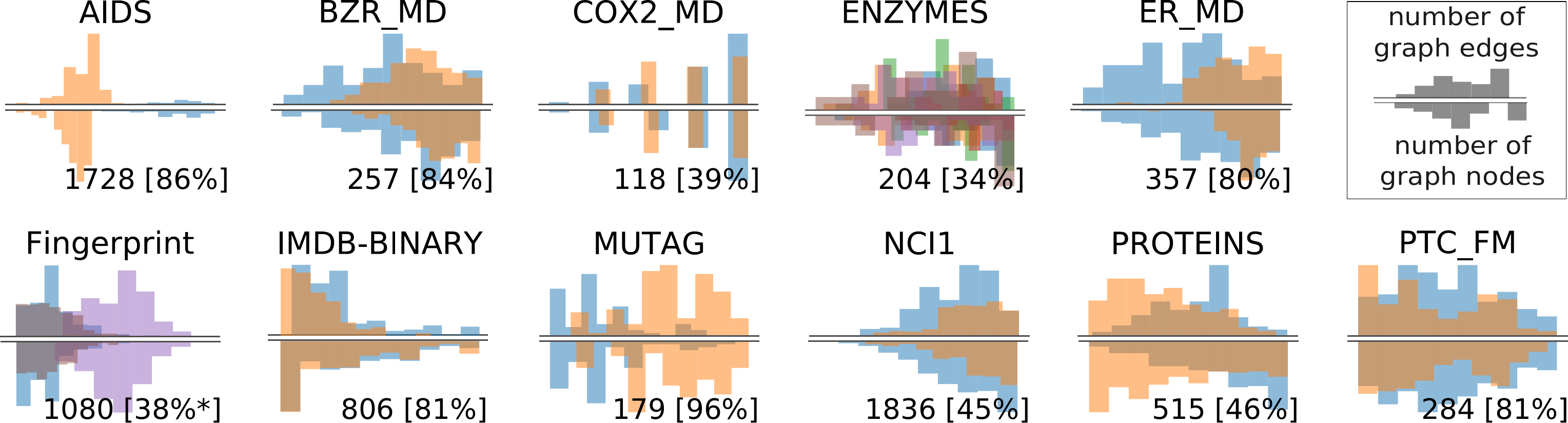}
\caption{Histograms of node and edge numbers of graphs in the benchmark datasets. The number of graphs as well as its percentage with respect to the original data are shown below each plot.}
\label{Fig:datasets}
\end{figure*}

Finally, we provide some numerical results to investigate the GBS graph kernel in practice. Benchmarks suggest that it is well competitive to standard ``classical'' graph kernels, at least in the hypothetical case of a perfect device. We furthermore show that displacement may improve classification accuracy by shifting weight into the higher-order orbits, and that orbits with photon numbers smaller or equal to $2$ contribute most to the result. 

\subsection{Benchmarking}

\begin{table*}
\centering
\def \arraystretch{1.2}
\begin{tabular}{lcccccccc}
\hline \hline
Dataset & GBS ($d_{0.0}$) & GBS ($d_{0.25}$) & GBS$^+$ ($d_{0.0}$) & GBS$^+$ ($d_{0.25}$) & GS & RW  & SM \\
\hline
AIDS &$99.60 \pm 0.05$ &$\mathbf{99.62} \pm 0.03$ &$99.58 \pm 00.06$ & $99.61 \pm 0.05$ & $98.44 \pm 0.09$& $56.95 \pm 7.99$ & $79.20 \pm 0.68$ \\
BZR\_MD &$62.73 \pm 0.71$&$62.13 \pm 1.44 $ &$62.01 \pm 1.43$ & $\mathbf{63.16} \pm 2.11 $ &$60.60 \pm 1.77$&$49.88 \pm 3.74$&$61.90 \pm 1.21$ \\
COX2\_MD &$44.98 \pm 1.80$& $ 50.11\pm 0.97$  &$57.84 \pm 4.04$ & $57.89 \pm 2.62 $ &$55.04 \pm 3.33 $&$57.72 \pm 3.26$& $\mathbf{66.94} \pm 1.22 $\\
ENZYMES &$22.29 \pm 1.60$ & $28.01 \pm 1.83$ &$25.72 \pm 2.60$ & $ \mathbf{40.42} \pm 2.02 $ &$35.87 \pm 2.19$&$21.13 \pm 1.91$ & $36.70 \pm 2.83$\\
ER\_MD &$70.36 \pm 0.78$& $70.41 \pm 0.47$ &$ 71.01 \pm 1.26 $ & $ \mathbf{71.05} \pm 0.83 $ &$65.65 \pm 1.06$&$68.75 \pm 0.53$& $68.21 \pm 0.99$\\
FINGERPRINT & $65.42 \pm 0.49$& $\textbf{65.85} \pm 0.36 $&$66.19 \pm 00.84$&$66.26 \pm 4.29$&$64.10 \pm 1.52$ &$47.69 \pm 0.21$& $47.14 \pm 0.62$\\
IMDB-BIN &$64.09 \pm 0.34$& $\textbf{68.71} \pm 0.59$&$ 68.14 \pm 0.71 $& $67.60 \pm 0.75 $&$68.37 \pm 0.62$ &$66.38 \pm 0.21$ & out of time$^*$ \\
MUTAG &$\mathbf{86.41} \pm 0.33$& $85.58 \pm 0.59$&$85.64 \pm 0.78 $&$84.46 \pm 0.44$&$81.08 \pm 0.93$&$83.02 \pm 1.08$&$83.14 \pm 0.24$\\
NCI1 &$\mathbf{63.61} \pm 0.00$& $62.79 \pm 0.00$ &$63.59 \pm 0.17 $&$63.11 \pm 0.93$& $49.96 \pm 3.27$ &$52.36 \pm 2.63$ &$51.36 \pm 1.88$ \\
PROTEINS &$\mathbf{66.88} \pm 0.22$& $66.14 \pm 0.48$&$65.73 \pm 0.69$&$66.16 \pm 0.76 $& $65.91 \pm 1.29$& $56.27 \pm 1.23$  & $63.03 \pm 0.84$\\
PTC\_FM &$53.84 \pm 0.96$& $52.45 \pm 1.78$ &$ 59.14 \pm 1.72$& $56.25 \pm 2.04 $& $\mathbf{59.48} \pm 1.95$& $51.97 \pm 2.68$ & $54.92 \pm 2.94 $ \\
\hline \hline
\end{tabular}
\caption{Mean test accuracy of the Support Vector Machine with different datasets and different graph kernels, with the standard deviation between $10$ repetitions of the double cross-validation. GS, RW, and SM are three standard classical graph kernels described in the text. GBS refers to the postprocessing strategy of associating orbit probabilities with features, while GBS$^+$ summarises some orbits to meta-orbits (see text).$^*$\textit{Runtime $>20$ days.} }
\label{Tbl:results}
\end{table*}

To benchmark the GBS feature map, we use a setup that has become a standard in testing graph kernels: A C-Support Vector Machine (SVM) with a precomputed kernel. The test accuracies in Table~\ref{Tbl:results} are obtained by running $10$ repeats of a double $10$-fold cross-validation. The inner fold extracts the best model by adjusting the $C$-parameter of the SVM -- which controls the penalty on misclassifications -- via grid search between values $[10^{-4}, 10^3]$, and the best model is then used to get the accuracy of the test set in the outer cross-validation loop. The GBS feature vectors were used in conjunction with a `rbf' kernel $\kappa_{\mathrm{rbf}}$. 

For the GBS graph kernel, we chose a gentle displacement of $d=0.25$ on every mode and $k=6$, leading to $30$-dimensional feature vectors. We used exact simulations based on the hafnian library \cite{bjorklund2018faster}. These are computationally very expensive, which is why we only consider small datasets.
Three classical graph kernels are benchmarked for comparison: The Graphlet Sampling kernel \cite{shervashidze2009efficient} (GS) with maximum graphlet size of $k=5$ and $5174$ samples drawn, the Random Walk kernel \cite{gartner2003graph} (RW) with fast computation and a geometric kernel type, and the Subgraph Matching kernel (SM) \cite{kriege2012subgraph}.
The three classical kernels were simulated using Python's \textit{grakel} library \cite{siglidis2018grakel}.\footnote{Experiments were run on IBM's cloud platform using four 2.8GHz Intel Xeon-IvyBridge Ex (E7-4890-V2-PentadecaCore) processors with 15 CPU cores each, as well as on Oak Ridge's Titan supercomputer.}

The datasets are taken from the repository of the Technical University of Dortmund \cite{KKMMN2016} (see Figure~\ref{Fig:datasets}). Preprocessing of the benchmarking datasets includes these three steps:
\begin{enumerate}
\item \textit{Graph selection}: Graphs which have less than $6$ or more than $25$ nodes are excluded to keep the feature vectors constant and to limit the time of simulations. The share of excluded graphs is displayed in Figure~(3) in the main paper, and ranges from $5\%$ to $55\%$.
\item \textit{Labels and attributes}: Potential node labels, node attributes and edge attributes are ignored. The edge labels in BZR\_MD, COX2\_MD, ER\_MD, MUTAG and PTC\_FM were translated to the following weights: $0$ - no chemical bond, $1$ -  single bond/double bond/triple bond/aromatic bond. 
The edge labels in AIDS where translated into the weights: 	$0$ - no edge $1$ - valence of zero, one or two.
In FINGERPRINT, only graphs of the three dominant classes $0,4,5$ were considered, since the other classes did not contain a sufficient number of samples after graph selection.
\item \textit{Rescaling}: The final (weighed or unweighed) adjacency matrix is divided by a normalization constant $c = 1/(\lambda^{\{G\}}_{\mathrm{max}} + 10^{-8}) $ that is slightly larger than the largest singular value $s^{\{G\}}_{\mathrm{max}} $ of any adjacency matrix in the dataset, as explained in Section~2.1 of the main paper.
\end{enumerate}
All datasets were chosen \textit{before} the first experiments were run, to avoid a post-selection bias in favour of the GBS kernel.

As Table~\ref{Tbl:results} shows, the GBS kernel performs well and outperforms the other methods visibly for MUTAG and NCI1, while still leading for AIDS, BZR\_MD, ER\_MD, FINGERPRINT and PROTEINS. Displacement increases the performance of the GBS kernel significantly for COX2\_MD, ENZYMES and IMDB-BIN, but not for other data sets. The GBS kernel does well on datasets where the distribution of node and edge numbers differs strongly between classes. However, we confirmed that excluding the `edge counting features' $[1,1,0..], [2,2,0..], \dots$ does not influence classification performance. While the graph size is considered by the GBS kernel, it seems to be only one of many properties that enters the notion of similarity. 

\subsection{Displacement and feature importance}

\begin{figure*}
\includegraphics[width=0.9\textwidth,valign=t]{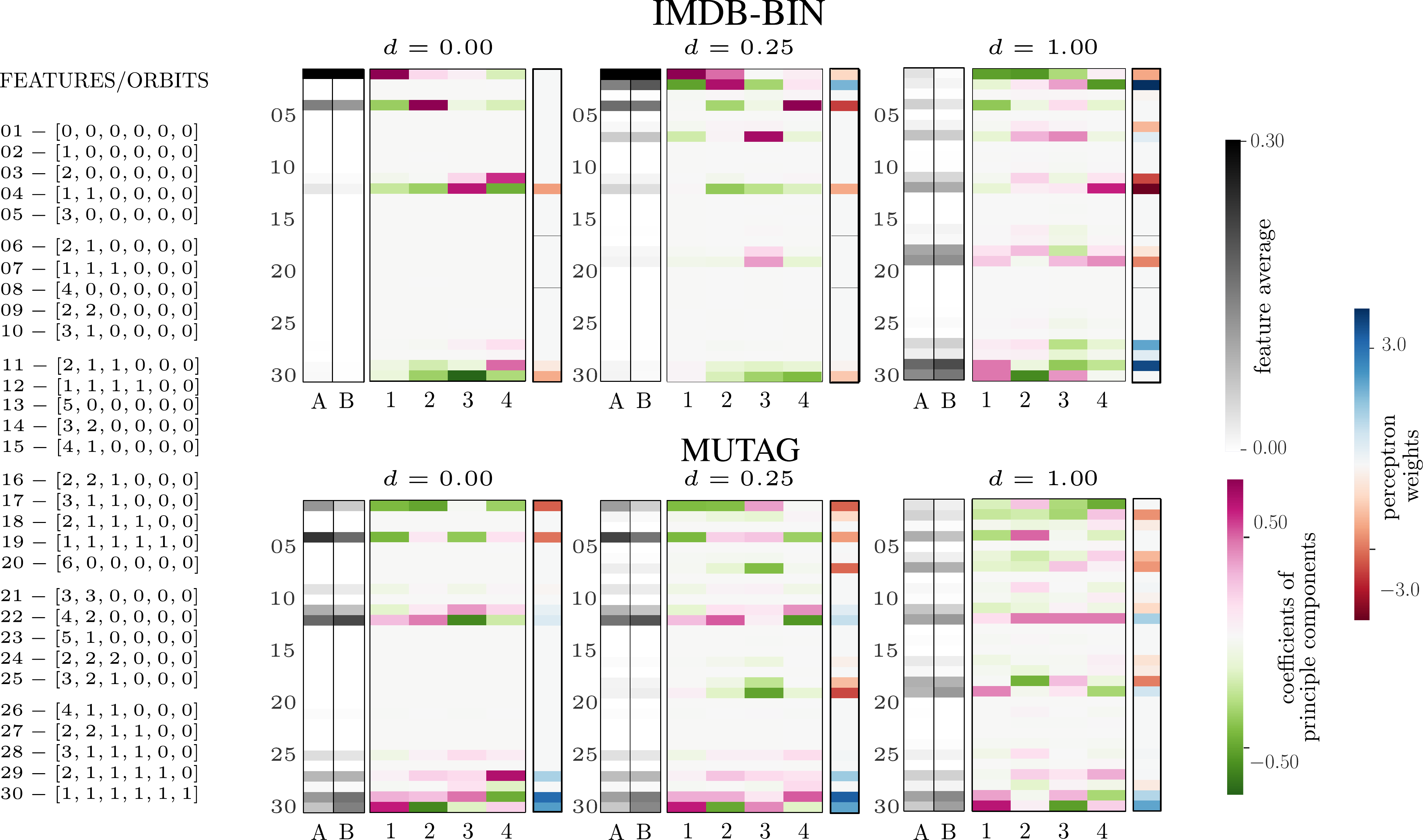}
\caption{Three measures for feature importance for IMDB-BINARY (top row) and MUTAG (bottom row) using $k=6$ and for $d=0, d=0.25$ and $d=1$. The $3 + 3$ heatmaps consist of three columns each. The leftmost column (gray color map) shows the average of each feature for the two different classes, here labeled $A$ and $B$. The center column shows the coefficients with which each feature contributes to the four first principal components in the PCA analysis. The third column shows the weights which a perceptron attributes to each feature when trained to classify the target labels. }
\label{Fig:import}
\end{figure*}

The hyperparameters of the GBS and GBS$^+$ graph kernels are the constant displacement $d$ which adiministered to each node, as well as the maximum photon number $k$. Since simulations restrict the value of $k$ at this stage, we focus on the effect of displacement, using the orbit-features (i.e., the GBS kernel). Displacement can change the similarity measure significantly. For example, comparing graphs of size $|V|=3$, one finds that the fully disconnected graph is closer to the fully connected graph than a graph with two edges for $d=1$, but vice versa for $d=0$. 

Figure~\ref{Fig:import} uses the example of IMDB-BIN and MUTAG to investigate the GBS$_{I}$ or ``orbit'' features for $d=0, d=0.25$ and $d=1$. 
The feature averages show that the general distribution of the feature vector is similar for both classes, but still visually distinguishable.\footnote{Standarization of the feature vectors to emphasize their mutual differences improved classification accuracy in some cases, but deteriorated it in others.} Consistent with the theory, increasing displacement shifts the features towards higher-order orbits, and populates features that are zero when $d=0$. 
Features associated with orbits $[1,1,0,...], [1,1,1,1,0...]$ and $[1,1,1,1,1,1]$, as well as $[2,1,0,...]$ and $[2,1,1,1,0,...]$ seem to be particularly important in the support of principal components, and get high weights when training a perceptron on the GBS features. Where displacement renders them nonzero, uneven orbits such as $[1,1,1,0,...], [1,1,1,1,1,0...]$ follow suit. During our investigations we confirmed that dropping features with high single-detector photon numbers did not have a huge influence on classification.
Consistent with the results from Table~\ref{Tbl:results}, MUTAG has `richer' features for $d=0$ than IMDB-BIN for classification with a perceptron, an advantage that IMDB-BIN equalizes with growing displacement. 

The feature analysis suggests that features related to subgraphs of all sizes (here $1$ to $6$) are important for the classification results, and that duplication of a single node in the subgraphs may be beneficial -- a feature that Graphlet Sampling kernels do not explore. The effect of displacement varies with the dataset, and $d$ should therefore be kept as a hyperparameter for model selection.

\section{Conclusion}

We proposed a new type of feature extraction strategy for graph-structured data based on the quantum technique of Gaussian Boson Sampling. We suggested that the success of the method is related to the fact that such a system samples from distributions that are related to useful graph properties. For classical machine learning, this method presents a potentially powerful extension to the gallery of graph kernels, each of which has strengths on certain data sets. For quantum machine learning, this proposes the first application of a ``quantum kernel''.

A lot of questions are still open for further investigation, for example regarding the role and interpretation of displacement, how GBS performs with weighted adjacency matrices, how node and edge labels can be considered, as well as whether the feature vectors are useful in combination with other methods such as neural networks. 
We expect that the rapid current development of numeric GBS samplers as well as quantum hardware will help answering these questions in the near future.

\section*{Acknowledgements}
We thank Christopher Morris and Nicolas Quesada for valuable advice, as well as the authors of Python's GraKel library. This research used resources of the Oak Ridge Leadership Computing Facility, which is a DOE Office of Science User Facility supported under Contract DE-AC05-00OR22725.


\begin{thebibliography}{50}%
\makeatletter
\providecommand \@ifxundefined [1]{%
 \@ifx{#1\undefined}
}%
\providecommand \@ifnum [1]{%
 \ifnum #1\expandafter \@firstoftwo
 \else \expandafter \@secondoftwo
 \fi
}%
\providecommand \@ifx [1]{%
 \ifx #1\expandafter \@firstoftwo
 \else \expandafter \@secondoftwo
 \fi
}%
\providecommand \natexlab [1]{#1}%
\providecommand \enquote  [1]{``#1''}%
\providecommand \bibnamefont  [1]{#1}%
\providecommand \bibfnamefont [1]{#1}%
\providecommand \citenamefont [1]{#1}%
\providecommand \href@noop [0]{\@secondoftwo}%
\providecommand \href [0]{\begingroup \@sanitize@url \@href}%
\providecommand \@href[1]{\@@startlink{#1}\@@href}%
\providecommand \@@href[1]{\endgroup#1\@@endlink}%
\providecommand \@sanitize@url [0]{\catcode `\\12\catcode `\$12\catcode
  `\&12\catcode `\#12\catcode `\^12\catcode `\_12\catcode `\%12\relax}%
\providecommand \@@startlink[1]{}%
\providecommand \@@endlink[0]{}%
\providecommand \url  [0]{\begingroup\@sanitize@url \@url }%
\providecommand \@url [1]{\endgroup\@href {#1}{\urlprefix }}%
\providecommand \urlprefix  [0]{URL }%
\providecommand \Eprint [0]{\href }%
\providecommand \doibase [0]{http://dx.doi.org/}%
\providecommand \selectlanguage [0]{\@gobble}%
\providecommand \bibinfo  [0]{\@secondoftwo}%
\providecommand \bibfield  [0]{\@secondoftwo}%
\providecommand \translation [1]{[#1]}%
\providecommand \BibitemOpen [0]{}%
\providecommand \bibitemStop [0]{}%
\providecommand \bibitemNoStop [0]{.\EOS\space}%
\providecommand \EOS [0]{\spacefactor3000\relax}%
\providecommand \BibitemShut  [1]{\csname bibitem#1\endcsname}%
\let\auto@bib@innerbib\@empty
\bibitem [{\citenamefont {Kobler}\ \emph {et~al.}(2012)\citenamefont {Kobler},
  \citenamefont {Sch{\"o}ning},\ and\ \citenamefont
  {Tor{\'a}n}}]{kobler2012graph}%
  \BibitemOpen
  \bibfield  {author} {\bibinfo {author} {\bibfnamefont {Johannes}\
  \bibnamefont {Kobler}}, \bibinfo {author} {\bibfnamefont {Uwe}\ \bibnamefont
  {Sch{\"o}ning}}, \ and\ \bibinfo {author} {\bibfnamefont {Jacobo}\
  \bibnamefont {Tor{\'a}n}},\ }\href@noop {} {\emph {\bibinfo {title} {The
  graph isomorphism problem: its structural complexity}}}\ (\bibinfo
  {publisher} {Springer Science \& Business Media},\ \bibinfo {year}
  {2012})\BibitemShut {NoStop}%
\bibitem [{\citenamefont {McKay}\ \emph {et~al.}(1981)\citenamefont {McKay}
  \emph {et~al.}}]{mckay1981practical}%
  \BibitemOpen
  \bibfield  {author} {\bibinfo {author} {\bibfnamefont {Brendan~D}\
  \bibnamefont {McKay}} \emph {et~al.},\ }\href@noop {} {\emph {\bibinfo
  {title} {Practical graph isomorphism}}}\ (\bibinfo  {publisher} {Department
  of Computer Science, Vanderbilt University Tennessee, USA},\ \bibinfo {year}
  {1981})\BibitemShut {NoStop}%
\bibitem [{\citenamefont {Ghosh}\ \emph {et~al.}(2018)\citenamefont {Ghosh},
  \citenamefont {Das}, \citenamefont {Gon{\c{c}}alves}, \citenamefont
  {Quaresma},\ and\ \citenamefont {Kundu}}]{ghosh2018journey}%
  \BibitemOpen
  \bibfield  {author} {\bibinfo {author} {\bibfnamefont {Swarnendu}\
  \bibnamefont {Ghosh}}, \bibinfo {author} {\bibfnamefont {Nibaran}\
  \bibnamefont {Das}}, \bibinfo {author} {\bibfnamefont {Teresa}\ \bibnamefont
  {Gon{\c{c}}alves}}, \bibinfo {author} {\bibfnamefont {Paulo}\ \bibnamefont
  {Quaresma}}, \ and\ \bibinfo {author} {\bibfnamefont {Mahantapas}\
  \bibnamefont {Kundu}},\ }\bibfield  {title} {\enquote {\bibinfo {title} {The
  journey of graph kernels through two decades},}\ }\href@noop {} {\bibfield
  {journal} {\bibinfo  {journal} {Computer Science Review}\ }\textbf {\bibinfo
  {volume} {27}},\ \bibinfo {pages} {88--111} (\bibinfo {year}
  {2018})}\BibitemShut {NoStop}%
\bibitem [{\citenamefont {Hamilton}\ \emph {et~al.}(2017)\citenamefont
  {Hamilton}, \citenamefont {Kruse}, \citenamefont {Sansoni}, \citenamefont
  {Barkhofen}, \citenamefont {Silberhorn},\ and\ \citenamefont
  {Jex}}]{hamilton2017gaussian}%
  \BibitemOpen
  \bibfield  {author} {\bibinfo {author} {\bibfnamefont {Craig~S}\ \bibnamefont
  {Hamilton}}, \bibinfo {author} {\bibfnamefont {Regina}\ \bibnamefont
  {Kruse}}, \bibinfo {author} {\bibfnamefont {Linda}\ \bibnamefont {Sansoni}},
  \bibinfo {author} {\bibfnamefont {Sonja}\ \bibnamefont {Barkhofen}}, \bibinfo
  {author} {\bibfnamefont {Christine}\ \bibnamefont {Silberhorn}}, \ and\
  \bibinfo {author} {\bibfnamefont {Igor}\ \bibnamefont {Jex}},\ }\bibfield
  {title} {\enquote {\bibinfo {title} {Gaussian boson sampling},}\ }\href@noop
  {} {\bibfield  {journal} {\bibinfo  {journal} {Physical review letters}\
  }\textbf {\bibinfo {volume} {119}},\ \bibinfo {pages} {170501} (\bibinfo
  {year} {2017})}\BibitemShut {NoStop}%
\bibitem [{\citenamefont {Lund}\ \emph {et~al.}(2014)\citenamefont {Lund},
  \citenamefont {Laing}, \citenamefont {Rahimi-Keshari}, \citenamefont
  {Rudolph}, \citenamefont {O’Brien},\ and\ \citenamefont
  {Ralph}}]{lund2014boson}%
  \BibitemOpen
  \bibfield  {author} {\bibinfo {author} {\bibfnamefont {AP}~\bibnamefont
  {Lund}}, \bibinfo {author} {\bibfnamefont {A}~\bibnamefont {Laing}}, \bibinfo
  {author} {\bibfnamefont {S}~\bibnamefont {Rahimi-Keshari}}, \bibinfo {author}
  {\bibfnamefont {T}~\bibnamefont {Rudolph}}, \bibinfo {author} {\bibfnamefont
  {Jeremy~L}\ \bibnamefont {O’Brien}}, \ and\ \bibinfo {author}
  {\bibfnamefont {TC}~\bibnamefont {Ralph}},\ }\bibfield  {title} {\enquote
  {\bibinfo {title} {Boson sampling from a gaussian state},}\ }\href@noop {}
  {\bibfield  {journal} {\bibinfo  {journal} {Physical review letters}\
  }\textbf {\bibinfo {volume} {113}},\ \bibinfo {pages} {100502} (\bibinfo
  {year} {2014})}\BibitemShut {NoStop}%
\bibitem [{\citenamefont {Kruse}\ \emph {et~al.}(2018)\citenamefont {Kruse},
  \citenamefont {Hamilton}, \citenamefont {Sansoni}, \citenamefont {Barkhofen},
  \citenamefont {Silberhorn},\ and\ \citenamefont {Jex}}]{kruse2018detailed}%
  \BibitemOpen
  \bibfield  {author} {\bibinfo {author} {\bibfnamefont {Regina}\ \bibnamefont
  {Kruse}}, \bibinfo {author} {\bibfnamefont {Craig~S}\ \bibnamefont
  {Hamilton}}, \bibinfo {author} {\bibfnamefont {Linda}\ \bibnamefont
  {Sansoni}}, \bibinfo {author} {\bibfnamefont {Sonja}\ \bibnamefont
  {Barkhofen}}, \bibinfo {author} {\bibfnamefont {Christine}\ \bibnamefont
  {Silberhorn}}, \ and\ \bibinfo {author} {\bibfnamefont {Igor}\ \bibnamefont
  {Jex}},\ }\bibfield  {title} {\enquote {\bibinfo {title} {{A detailed study
  of Gaussian Boson Sampling}},}\ }\href@noop {} {\bibfield  {journal}
  {\bibinfo  {journal} {arXiv preprint arXiv:1801.07488}\ } (\bibinfo {year}
  {2018})}\BibitemShut {NoStop}%
\bibitem [{\citenamefont {Tillmann}\ \emph {et~al.}(2013)\citenamefont
  {Tillmann}, \citenamefont {Daki{\'c}}, \citenamefont {Heilmann},
  \citenamefont {Nolte}, \citenamefont {Szameit},\ and\ \citenamefont
  {Walther}}]{tillmann2013experimental}%
  \BibitemOpen
  \bibfield  {author} {\bibinfo {author} {\bibfnamefont {Max}\ \bibnamefont
  {Tillmann}}, \bibinfo {author} {\bibfnamefont {Borivoje}\ \bibnamefont
  {Daki{\'c}}}, \bibinfo {author} {\bibfnamefont {Ren{\'e}}\ \bibnamefont
  {Heilmann}}, \bibinfo {author} {\bibfnamefont {Stefan}\ \bibnamefont
  {Nolte}}, \bibinfo {author} {\bibfnamefont {Alexander}\ \bibnamefont
  {Szameit}}, \ and\ \bibinfo {author} {\bibfnamefont {Philip}\ \bibnamefont
  {Walther}},\ }\bibfield  {title} {\enquote {\bibinfo {title} {Experimental
  boson sampling},}\ }\href@noop {} {\bibfield  {journal} {\bibinfo  {journal}
  {Nature Photonics}\ }\textbf {\bibinfo {volume} {7}},\ \bibinfo {pages} {540}
  (\bibinfo {year} {2013})}\BibitemShut {NoStop}%
\bibitem [{\citenamefont {Broome}\ \emph {et~al.}(2013)\citenamefont {Broome},
  \citenamefont {Fedrizzi}, \citenamefont {Rahimi-Keshari}, \citenamefont
  {Dove}, \citenamefont {Aaronson}, \citenamefont {Ralph},\ and\ \citenamefont
  {White}}]{broome2013photonic}%
  \BibitemOpen
  \bibfield  {author} {\bibinfo {author} {\bibfnamefont {Matthew~A}\
  \bibnamefont {Broome}}, \bibinfo {author} {\bibfnamefont {Alessandro}\
  \bibnamefont {Fedrizzi}}, \bibinfo {author} {\bibfnamefont {Saleh}\
  \bibnamefont {Rahimi-Keshari}}, \bibinfo {author} {\bibfnamefont {Justin}\
  \bibnamefont {Dove}}, \bibinfo {author} {\bibfnamefont {Scott}\ \bibnamefont
  {Aaronson}}, \bibinfo {author} {\bibfnamefont {Timothy~C}\ \bibnamefont
  {Ralph}}, \ and\ \bibinfo {author} {\bibfnamefont {Andrew~G}\ \bibnamefont
  {White}},\ }\bibfield  {title} {\enquote {\bibinfo {title} {Photonic boson
  sampling in a tunable circuit},}\ }\href@noop {} {\bibfield  {journal}
  {\bibinfo  {journal} {Science}\ }\textbf {\bibinfo {volume} {339}},\ \bibinfo
  {pages} {794--798} (\bibinfo {year} {2013})}\BibitemShut {NoStop}%
\bibitem [{\citenamefont {Aaronson}\ and\ \citenamefont
  {Arkhipov}(2011)}]{aaronson2011computational}%
  \BibitemOpen
  \bibfield  {author} {\bibinfo {author} {\bibfnamefont {Scott}\ \bibnamefont
  {Aaronson}}\ and\ \bibinfo {author} {\bibfnamefont {Alex}\ \bibnamefont
  {Arkhipov}},\ }\bibfield  {title} {\enquote {\bibinfo {title} {The
  computational complexity of linear optics},}\ }in\ \href@noop {} {\emph
  {\bibinfo {booktitle} {Proceedings of the forty-third annual ACM symposium on
  Theory of computing}}}\ (\bibinfo {organization} {ACM},\ \bibinfo {year}
  {2011})\ pp.\ \bibinfo {pages} {333--342}\BibitemShut {NoStop}%
\bibitem [{\citenamefont {Br{\'a}dler}\ \emph {et~al.}(2018)\citenamefont
  {Br{\'a}dler}, \citenamefont {Dallaire-Demers}, \citenamefont {Rebentrost},
  \citenamefont {Su},\ and\ \citenamefont {Weedbrook}}]{bradler2018gaussian}%
  \BibitemOpen
  \bibfield  {author} {\bibinfo {author} {\bibfnamefont {Kamil}\ \bibnamefont
  {Br{\'a}dler}}, \bibinfo {author} {\bibfnamefont {Pierre-Luc}\ \bibnamefont
  {Dallaire-Demers}}, \bibinfo {author} {\bibfnamefont {Patrick}\ \bibnamefont
  {Rebentrost}}, \bibinfo {author} {\bibfnamefont {Daiqin}\ \bibnamefont {Su}},
  \ and\ \bibinfo {author} {\bibfnamefont {Christian}\ \bibnamefont
  {Weedbrook}},\ }\bibfield  {title} {\enquote {\bibinfo {title} {{Gaussian
  boson sampling for perfect matchings of arbitrary graphs}},}\ }\href@noop {}
  {\bibfield  {journal} {\bibinfo  {journal} {Physical Review A}\ }\textbf
  {\bibinfo {volume} {98}},\ \bibinfo {pages} {032310} (\bibinfo {year}
  {2018})}\BibitemShut {NoStop}%
\bibitem [{\citenamefont {Br\'adler}\ \emph {et~al.}(2018)\citenamefont
  {Br\'adler}, \citenamefont {Friedland}, \citenamefont {Izaac}, \citenamefont
  {Killoran},\ and\ \citenamefont {Su}}]{bradler2018graph}%
  \BibitemOpen
  \bibfield  {author} {\bibinfo {author} {\bibfnamefont {Kamil}\ \bibnamefont
  {Br\'adler}}, \bibinfo {author} {\bibfnamefont {Shmuel}\ \bibnamefont
  {Friedland}}, \bibinfo {author} {\bibfnamefont {Josh}\ \bibnamefont {Izaac}},
  \bibinfo {author} {\bibfnamefont {Nathan}\ \bibnamefont {Killoran}}, \ and\
  \bibinfo {author} {\bibfnamefont {Daiqin}\ \bibnamefont {Su}},\ }\bibfield
  {title} {\enquote {\bibinfo {title} {{Graph isomorphism and Gaussian boson
  sampling}},}\ }\href@noop {} {\bibfield  {journal} {\bibinfo  {journal}
  {arXiv preprint arXiv:1810.10644}\ } (\bibinfo {year} {2018})}\BibitemShut
  {NoStop}%
\bibitem [{\citenamefont {Zhang}\ \emph {et~al.}(2018)\citenamefont {Zhang},
  \citenamefont {Yin}, \citenamefont {Zhu},\ and\ \citenamefont
  {Zhang}}]{zhang2018network}%
  \BibitemOpen
  \bibfield  {author} {\bibinfo {author} {\bibfnamefont {Daokun}\ \bibnamefont
  {Zhang}}, \bibinfo {author} {\bibfnamefont {Jie}\ \bibnamefont {Yin}},
  \bibinfo {author} {\bibfnamefont {Xingquan}\ \bibnamefont {Zhu}}, \ and\
  \bibinfo {author} {\bibfnamefont {Chengqi}\ \bibnamefont {Zhang}},\
  }\bibfield  {title} {\enquote {\bibinfo {title} {Network representation
  learning: A survey},}\ }\href@noop {} {\bibfield  {journal} {\bibinfo
  {journal} {IEEE transactions on Big Data}\ } (\bibinfo {year}
  {2018})}\BibitemShut {NoStop}%
\bibitem [{\citenamefont {Goyal}\ and\ \citenamefont
  {Ferrara}(2018)}]{goyal2018graph}%
  \BibitemOpen
  \bibfield  {author} {\bibinfo {author} {\bibfnamefont {Palash}\ \bibnamefont
  {Goyal}}\ and\ \bibinfo {author} {\bibfnamefont {Emilio}\ \bibnamefont
  {Ferrara}},\ }\bibfield  {title} {\enquote {\bibinfo {title} {Graph embedding
  techniques, applications, and performance: A survey},}\ }\href@noop {}
  {\bibfield  {journal} {\bibinfo  {journal} {Knowledge-Based Systems}\
  }\textbf {\bibinfo {volume} {151}},\ \bibinfo {pages} {78--94} (\bibinfo
  {year} {2018})}\BibitemShut {NoStop}%
\bibitem [{\citenamefont {Grover}\ and\ \citenamefont
  {Leskovec}(2016)}]{grover2016node2vec}%
  \BibitemOpen
  \bibfield  {author} {\bibinfo {author} {\bibfnamefont {Aditya}\ \bibnamefont
  {Grover}}\ and\ \bibinfo {author} {\bibfnamefont {Jure}\ \bibnamefont
  {Leskovec}},\ }\bibfield  {title} {\enquote {\bibinfo {title} {node2vec:
  Scalable feature learning for networks},}\ }in\ \href@noop {} {\emph
  {\bibinfo {booktitle} {Proceedings of the 22nd ACM SIGKDD international
  conference on Knowledge discovery and data mining}}}\ (\bibinfo
  {organization} {ACM},\ \bibinfo {year} {2016})\ pp.\ \bibinfo {pages}
  {855--864}\BibitemShut {NoStop}%
\bibitem [{\citenamefont {Kriege}\ \emph {et~al.}(2014)\citenamefont {Kriege},
  \citenamefont {Neumann}, \citenamefont {Kersting},\ and\ \citenamefont
  {Mutzel}}]{kriege2014explicit}%
  \BibitemOpen
  \bibfield  {author} {\bibinfo {author} {\bibfnamefont {Nils}\ \bibnamefont
  {Kriege}}, \bibinfo {author} {\bibfnamefont {Marion}\ \bibnamefont
  {Neumann}}, \bibinfo {author} {\bibfnamefont {Kristian}\ \bibnamefont
  {Kersting}}, \ and\ \bibinfo {author} {\bibfnamefont {Petra}\ \bibnamefont
  {Mutzel}},\ }\bibfield  {title} {\enquote {\bibinfo {title} {Explicit versus
  implicit graph feature maps: A computational phase transition for walk
  kernels},}\ }in\ \href@noop {} {\emph {\bibinfo {booktitle} {Data Mining
  (ICDM), 2014 IEEE International Conference on}}}\ (\bibinfo {organization}
  {IEEE},\ \bibinfo {year} {2014})\ pp.\ \bibinfo {pages}
  {881--886}\BibitemShut {NoStop}%
\bibitem [{\citenamefont {Schuld}\ and\ \citenamefont
  {Killoran}(2019)}]{schuld2019quantum}%
  \BibitemOpen
  \bibfield  {author} {\bibinfo {author} {\bibfnamefont {Maria}\ \bibnamefont
  {Schuld}}\ and\ \bibinfo {author} {\bibfnamefont {Nathan}\ \bibnamefont
  {Killoran}},\ }\bibfield  {title} {\enquote {\bibinfo {title} {{Quantum
  machine learning in feature Hilbert spaces}},}\ }\href@noop {} {\bibfield
  {journal} {\bibinfo  {journal} {Physical review letters}\ }\textbf {\bibinfo
  {volume} {122}},\ \bibinfo {pages} {040504} (\bibinfo {year}
  {2019})}\BibitemShut {NoStop}%
\bibitem [{\citenamefont {Havl{\'\i}{\v{c}}ek}\ \emph
  {et~al.}(2019)\citenamefont {Havl{\'\i}{\v{c}}ek}, \citenamefont
  {C{\'o}rcoles}, \citenamefont {Temme}, \citenamefont {Harrow}, \citenamefont
  {Kandala}, \citenamefont {Chow},\ and\ \citenamefont
  {Gambetta}}]{havlivcek2019supervised}%
  \BibitemOpen
  \bibfield  {author} {\bibinfo {author} {\bibfnamefont {Vojt{\v{e}}ch}\
  \bibnamefont {Havl{\'\i}{\v{c}}ek}}, \bibinfo {author} {\bibfnamefont
  {Antonio~D}\ \bibnamefont {C{\'o}rcoles}}, \bibinfo {author} {\bibfnamefont
  {Kristan}\ \bibnamefont {Temme}}, \bibinfo {author} {\bibfnamefont {Aram~W}\
  \bibnamefont {Harrow}}, \bibinfo {author} {\bibfnamefont {Abhinav}\
  \bibnamefont {Kandala}}, \bibinfo {author} {\bibfnamefont {Jerry~M}\
  \bibnamefont {Chow}}, \ and\ \bibinfo {author} {\bibfnamefont {Jay~M}\
  \bibnamefont {Gambetta}},\ }\bibfield  {title} {\enquote {\bibinfo {title}
  {Supervised learning with quantum-enhanced feature spaces},}\ }\href@noop {}
  {\bibfield  {journal} {\bibinfo  {journal} {Nature}\ }\textbf {\bibinfo
  {volume} {567}},\ \bibinfo {pages} {209} (\bibinfo {year}
  {2019})}\BibitemShut {NoStop}%
\bibitem [{\citenamefont {Scholkopf}\ and\ \citenamefont
  {Smola}(2001)}]{scholkopf2001learning}%
  \BibitemOpen
  \bibfield  {author} {\bibinfo {author} {\bibfnamefont {Bernhard}\
  \bibnamefont {Scholkopf}}\ and\ \bibinfo {author} {\bibfnamefont
  {Alexander~J}\ \bibnamefont {Smola}},\ }\href@noop {} {\emph {\bibinfo
  {title} {Learning with kernels: support vector machines, regularization,
  optimization, and beyond}}}\ (\bibinfo  {publisher} {MIT press},\ \bibinfo
  {year} {2001})\BibitemShut {NoStop}%
\bibitem [{\citenamefont {Shervashidze}\ \emph {et~al.}(2009)\citenamefont
  {Shervashidze}, \citenamefont {Vishwanathan}, \citenamefont {Petri},
  \citenamefont {Mehlhorn},\ and\ \citenamefont
  {Borgwardt}}]{shervashidze2009efficient}%
  \BibitemOpen
  \bibfield  {author} {\bibinfo {author} {\bibfnamefont {Nino}\ \bibnamefont
  {Shervashidze}}, \bibinfo {author} {\bibfnamefont {SVN}\ \bibnamefont
  {Vishwanathan}}, \bibinfo {author} {\bibfnamefont {Tobias}\ \bibnamefont
  {Petri}}, \bibinfo {author} {\bibfnamefont {Kurt}\ \bibnamefont {Mehlhorn}},
  \ and\ \bibinfo {author} {\bibfnamefont {Karsten}\ \bibnamefont
  {Borgwardt}},\ }\bibfield  {title} {\enquote {\bibinfo {title} {Efficient
  graphlet kernels for large graph comparison},}\ }in\ \href@noop {} {\emph
  {\bibinfo {booktitle} {Artificial Intelligence and Statistics}}}\ (\bibinfo
  {year} {2009})\ pp.\ \bibinfo {pages} {488--495}\BibitemShut {NoStop}%
\bibitem [{\citenamefont {Weedbrook}\ \emph {et~al.}(2012)\citenamefont
  {Weedbrook}, \citenamefont {Pirandola}, \citenamefont
  {Garc{\'\i}a-Patr{\'o}n}, \citenamefont {Cerf}, \citenamefont {Ralph},
  \citenamefont {Shapiro},\ and\ \citenamefont
  {Lloyd}}]{weedbrook2012gaussian}%
  \BibitemOpen
  \bibfield  {author} {\bibinfo {author} {\bibfnamefont {Christian}\
  \bibnamefont {Weedbrook}}, \bibinfo {author} {\bibfnamefont {Stefano}\
  \bibnamefont {Pirandola}}, \bibinfo {author} {\bibfnamefont {Ra{\'u}l}\
  \bibnamefont {Garc{\'\i}a-Patr{\'o}n}}, \bibinfo {author} {\bibfnamefont
  {Nicolas~J}\ \bibnamefont {Cerf}}, \bibinfo {author} {\bibfnamefont
  {Timothy~C}\ \bibnamefont {Ralph}}, \bibinfo {author} {\bibfnamefont
  {Jeffrey~H}\ \bibnamefont {Shapiro}}, \ and\ \bibinfo {author} {\bibfnamefont
  {Seth}\ \bibnamefont {Lloyd}},\ }\bibfield  {title} {\enquote {\bibinfo
  {title} {Gaussian quantum information},}\ }\href@noop {} {\bibfield
  {journal} {\bibinfo  {journal} {Reviews of Modern Physics}\ }\textbf
  {\bibinfo {volume} {84}},\ \bibinfo {pages} {621} (\bibinfo {year}
  {2012})}\BibitemShut {NoStop}%
\bibitem [{Note1()}]{Note1}%
  \BibitemOpen
  \bibinfo {note} {As long as it fulfills the above inequality, $c$ can be
  treated as a hyperparameter of the feature map, which may also be influenced
  by hardware constraints since it relates ultimately to the amount of
  squeezing required.}\BibitemShut {Stop}%
\bibitem [{\citenamefont {Valiant}(1979)}]{valiant1979complexity}%
  \BibitemOpen
  \bibfield  {author} {\bibinfo {author} {\bibfnamefont {Leslie~G}\
  \bibnamefont {Valiant}},\ }\bibfield  {title} {\enquote {\bibinfo {title}
  {The complexity of computing the permanent},}\ }\href@noop {} {\bibfield
  {journal} {\bibinfo  {journal} {Theoretical computer science}\ }\textbf
  {\bibinfo {volume} {8}},\ \bibinfo {pages} {189--201} (\bibinfo {year}
  {1979})}\BibitemShut {NoStop}%
\bibitem [{\citenamefont {Barvinok}(2016)}]{barvinok2016approximating}%
  \BibitemOpen
  \bibfield  {author} {\bibinfo {author} {\bibfnamefont {Alexander}\
  \bibnamefont {Barvinok}},\ }\bibfield  {title} {\enquote {\bibinfo {title}
  {Approximating permanents and hafnians},}\ }\href@noop {} {\bibfield
  {journal} {\bibinfo  {journal} {arXiv preprint arXiv:1601.07518}\ } (\bibinfo
  {year} {2016})}\BibitemShut {NoStop}%
\bibitem [{\citenamefont {Rudelson}\ \emph {et~al.}(2016)\citenamefont
  {Rudelson}, \citenamefont {Samorodnitsky}, \citenamefont {Zeitouni} \emph
  {et~al.}}]{rudelson2016hafnians}%
  \BibitemOpen
  \bibfield  {author} {\bibinfo {author} {\bibfnamefont {Mark}\ \bibnamefont
  {Rudelson}}, \bibinfo {author} {\bibfnamefont {Alex}\ \bibnamefont
  {Samorodnitsky}}, \bibinfo {author} {\bibfnamefont {Ofer}\ \bibnamefont
  {Zeitouni}},  \emph {et~al.},\ }\bibfield  {title} {\enquote {\bibinfo
  {title} {Hafnians, perfect matchings and gaussian matrices},}\ }\href@noop {}
  {\bibfield  {journal} {\bibinfo  {journal} {The Annals of Probability}\
  }\textbf {\bibinfo {volume} {44}},\ \bibinfo {pages} {2858--2888} (\bibinfo
  {year} {2016})}\BibitemShut {NoStop}%
\bibitem [{Note2()}]{Note2}%
  \BibitemOpen
  \bibinfo {note} {To derive Eq.~(\ref {Eq:displ}) from the analysis in \cite
  {kruse2018detailed}, one uses the fact that for $\protect \mathaccentV
  {tilde}07E{A}$ being a direct sum $A \oplus A$, the index set $i_1,\protect
  \dots ,i_{2n} \in \protect \mathcal {I}_{2M}$ can be divided into two index
  sets: $j_1,\protect \dots ,j_s$ which contains all $s$ indices from the
  `first subspace' (i.e., the first $M$ dimensions) of $\protect \mathaccentV
  {tilde}07E{A}$, and $k_1,\protect \dots , k_{s'}$ containing the $s'$ indices
  from the `second subspace', with $s+s' = 2n$. The fact that $\protect \mathrm
  {haf}(A \oplus B) = \protect \mathrm {haf}(A)\protect \mathrm {haf}(B)$,
  allows us to express the Hafnian of reduced versions of $\protect
  \mathaccentV {tilde}07E{A}_{\protect \mathbf {n}} $ as a product of reduced
  versions of matrix $\protect \mathaccentV {tilde}07E{A}_{\protect \mathbf
  {n}} $, $$ \protect \mathrm {haf}(\protect \mathaccentV
  {tilde}07E{A}_{\protect \mathbf {n}- \protect \{i_1,\protect \dots
  ,i_{2n}\protect \}}) = \protect \mathrm {haf}(A_{\protect \mathbf {n}-
  \protect \{j_1,\protect \dots ,j_{s}\protect \}}) \protect \mathrm
  {haf}(A_{\protect \mathbf {n}- \protect \{k_1,\protect \dots ,k_{s'}\protect
  \}}). $$}\BibitemShut {NoStop}%
\bibitem [{\citenamefont {Quesada}(2019)}]{quesada2019franck}%
  \BibitemOpen
  \bibfield  {author} {\bibinfo {author} {\bibfnamefont {Nicol{\'a}s}\
  \bibnamefont {Quesada}},\ }\bibfield  {title} {\enquote {\bibinfo {title}
  {Franck-condon factors by counting perfect matchings of graphs with loops},}\
  }\href@noop {} {\bibfield  {journal} {\bibinfo  {journal} {The Journal of
  chemical physics}\ }\textbf {\bibinfo {volume} {150}},\ \bibinfo {pages}
  {164113} (\bibinfo {year} {2019})}\BibitemShut {NoStop}%
\bibitem [{Note3()}]{Note3}%
  \BibitemOpen
  \bibinfo {note} {See also A000070 in the Online Encyclopedia of Integer
  Sequences, \protect \url {https://oeis.org/A000070}.}\BibitemShut {Stop}%
\bibitem [{Note4()}]{Note4}%
  \BibitemOpen
  \bibinfo {note} {The energy of a Gaussian quantum state, and hence the
  average photon number, is determined by the squeezing and displacement
  operations.}\BibitemShut {Stop}%
\bibitem [{\citenamefont {Vaidya}\ \emph {et~al.}(2019)\citenamefont {Vaidya},
  \citenamefont {Morrison}, \citenamefont {Helt}, \citenamefont
  {Shahrokhshahi}, \citenamefont {Mahler}, \citenamefont {Collins},
  \citenamefont {Tan}, \citenamefont {Lavoie}, \citenamefont {Repingon},
  \citenamefont {Menotti} \emph {et~al.}}]{vaidya2019broadband}%
  \BibitemOpen
  \bibfield  {author} {\bibinfo {author} {\bibfnamefont {VD}~\bibnamefont
  {Vaidya}}, \bibinfo {author} {\bibfnamefont {B}~\bibnamefont {Morrison}},
  \bibinfo {author} {\bibfnamefont {LG}~\bibnamefont {Helt}}, \bibinfo {author}
  {\bibfnamefont {R}~\bibnamefont {Shahrokhshahi}}, \bibinfo {author}
  {\bibfnamefont {DH}~\bibnamefont {Mahler}}, \bibinfo {author} {\bibfnamefont
  {MJ}~\bibnamefont {Collins}}, \bibinfo {author} {\bibfnamefont
  {K}~\bibnamefont {Tan}}, \bibinfo {author} {\bibfnamefont {J}~\bibnamefont
  {Lavoie}}, \bibinfo {author} {\bibfnamefont {A}~\bibnamefont {Repingon}},
  \bibinfo {author} {\bibfnamefont {M}~\bibnamefont {Menotti}},  \emph
  {et~al.},\ }\bibfield  {title} {\enquote {\bibinfo {title} {Broadband
  quadrature-squeezed vacuum and nonclassical photon number correlations from a
  nanophotonic device},}\ }\href@noop {} {\bibfield  {journal} {\bibinfo
  {journal} {arXiv preprint arXiv:1904.07833}\ } (\bibinfo {year}
  {2019})}\BibitemShut {NoStop}%
\bibitem [{\citenamefont {{Farrell}}(1979)}]{farrell1979an}%
  \BibitemOpen
  \bibfield  {author} {\bibinfo {author} {\bibfnamefont {E.J}\ \bibnamefont
  {{Farrell}}},\ }\bibfield  {title} {\enquote {\bibinfo {title} {An
  introduction to matching polynomials},}\ }\href@noop {} {\bibfield  {journal}
  {\bibinfo  {journal} {Journal of Combinatorial Theory, Series B}\ }\textbf
  {\bibinfo {volume} {27}},\ \bibinfo {pages} {75--86} (\bibinfo {year}
  {1979})}\BibitemShut {NoStop}%
\bibitem [{\citenamefont {{Godsil}}\ and\ \citenamefont
  {{Gutman}}(1981)}]{godsil1981on}%
  \BibitemOpen
  \bibfield  {author} {\bibinfo {author} {\bibfnamefont {Chris~D.}\
  \bibnamefont {{Godsil}}}\ and\ \bibinfo {author} {\bibfnamefont {Ivan}\
  \bibnamefont {{Gutman}}},\ }\bibfield  {title} {\enquote {\bibinfo {title}
  {On the theory of the matching polynomial},}\ }\href@noop {} {\bibfield
  {journal} {\bibinfo  {journal} {Journal of Graph Theory}\ }\textbf {\bibinfo
  {volume} {5}},\ \bibinfo {pages} {137--144} (\bibinfo {year}
  {1981})}\BibitemShut {NoStop}%
\bibitem [{\citenamefont {Heilmann}\ and\ \citenamefont
  {Lieb}(1972)}]{heilmann1972theory}%
  \BibitemOpen
  \bibfield  {author} {\bibinfo {author} {\bibfnamefont {Ole~J}\ \bibnamefont
  {Heilmann}}\ and\ \bibinfo {author} {\bibfnamefont {Elliott~H}\ \bibnamefont
  {Lieb}},\ }\bibfield  {title} {\enquote {\bibinfo {title} {Theory of
  monomer-dimer systems},}\ }in\ \href@noop {} {\emph {\bibinfo {booktitle}
  {Statistical Mechanics}}}\ (\bibinfo  {publisher} {Springer},\ \bibinfo
  {year} {1972})\ pp.\ \bibinfo {pages} {45--87}\BibitemShut {NoStop}%
\bibitem [{\citenamefont {Averbouch}\ \emph {et~al.}(2008)\citenamefont
  {Averbouch}, \citenamefont {Godlin},\ and\ \citenamefont {Makowsky}}]{eep}%
  \BibitemOpen
  \bibfield  {author} {\bibinfo {author} {\bibfnamefont {Ilia}\ \bibnamefont
  {Averbouch}}, \bibinfo {author} {\bibfnamefont {Benny}\ \bibnamefont
  {Godlin}}, \ and\ \bibinfo {author} {\bibfnamefont {Johann~A.}\ \bibnamefont
  {Makowsky}},\ }\bibfield  {title} {\enquote {\bibinfo {title} {A most general
  edge elimination polynomial},}\ }in\ \href@noop {} {\emph {\bibinfo
  {booktitle} {Graph-Theoretic Concepts in Computer Science}}},\ \bibinfo
  {editor} {edited by\ \bibinfo {editor} {\bibfnamefont {Hajo}\ \bibnamefont
  {Broersma}}, \bibinfo {editor} {\bibfnamefont {Thomas}\ \bibnamefont
  {Erlebach}}, \bibinfo {editor} {\bibfnamefont {Tom}\ \bibnamefont
  {Friedetzky}}, \ and\ \bibinfo {editor} {\bibfnamefont {Daniel}\ \bibnamefont
  {Paulusma}}}\ (\bibinfo  {publisher} {Springer Berlin Heidelberg},\ \bibinfo
  {address} {Berlin, Heidelberg},\ \bibinfo {year} {2008})\ pp.\ \bibinfo
  {pages} {31--42}\BibitemShut {NoStop}%
\bibitem [{\citenamefont {Godsil}(1981)}]{godsil1981matchings}%
  \BibitemOpen
  \bibfield  {author} {\bibinfo {author} {\bibfnamefont {Christopher~David}\
  \bibnamefont {Godsil}},\ }\bibfield  {title} {\enquote {\bibinfo {title}
  {Matchings and walks in graphs},}\ }\href@noop {} {\bibfield  {journal}
  {\bibinfo  {journal} {Journal of Graph Theory}\ }\textbf {\bibinfo {volume}
  {5}},\ \bibinfo {pages} {285--297} (\bibinfo {year} {1981})}\BibitemShut
  {NoStop}%
\bibitem [{\citenamefont {Cvetkovic}\ \emph {et~al.}(1988)\citenamefont
  {Cvetkovic}, \citenamefont {Doob}, \citenamefont {Gutman},\ and\
  \citenamefont {Torga{\v{s}}ev}}]{cvetkovic1988recent}%
  \BibitemOpen
  \bibfield  {author} {\bibinfo {author} {\bibfnamefont {Dragos~M}\
  \bibnamefont {Cvetkovic}}, \bibinfo {author} {\bibfnamefont {Michael}\
  \bibnamefont {Doob}}, \bibinfo {author} {\bibfnamefont {Ivan}\ \bibnamefont
  {Gutman}}, \ and\ \bibinfo {author} {\bibfnamefont {Aleksandar}\ \bibnamefont
  {Torga{\v{s}}ev}},\ }\href@noop {} {\emph {\bibinfo {title} {Recent results
  in the theory of graph spectra}}},\ Vol.~\bibinfo {volume} {36}\ (\bibinfo
  {publisher} {Elsevier},\ \bibinfo {year} {1988})\BibitemShut {NoStop}%
\bibitem [{\citenamefont {Shi}\ \emph {et~al.}(2016)\citenamefont {Shi},
  \citenamefont {Dehmer}, \citenamefont {Li},\ and\ \citenamefont
  {Gutman}}]{shi2016graph}%
  \BibitemOpen
  \bibfield  {author} {\bibinfo {author} {\bibfnamefont {Yongtang}\
  \bibnamefont {Shi}}, \bibinfo {author} {\bibfnamefont {Matthias}\
  \bibnamefont {Dehmer}}, \bibinfo {author} {\bibfnamefont {Xueliang}\
  \bibnamefont {Li}}, \ and\ \bibinfo {author} {\bibfnamefont {Ivan}\
  \bibnamefont {Gutman}},\ }\href@noop {} {\emph {\bibinfo {title} {Graph
  Polynomials}}}\ (\bibinfo  {publisher} {Chapman and Hall/CRC},\ \bibinfo
  {year} {2016})\BibitemShut {NoStop}%
\bibitem [{\citenamefont {Godsil}(1993)}]{godsilAlgComb}%
  \BibitemOpen
  \bibfield  {author} {\bibinfo {author} {\bibfnamefont {Chris}\ \bibnamefont
  {Godsil}},\ }\href@noop {} {\emph {\bibinfo {title} {Algebraic
  Combinatorics}}}\ (\bibinfo  {publisher} {Chapman Hall Crc Mathematics
  Series},\ \bibinfo {year} {1993})\BibitemShut {NoStop}%
\bibitem [{\citenamefont {Lass}(2004)}]{lass2004matching}%
  \BibitemOpen
  \bibfield  {author} {\bibinfo {author} {\bibfnamefont {Bodo}\ \bibnamefont
  {Lass}},\ }\bibfield  {title} {\enquote {\bibinfo {title} {Matching
  polynomials and duality},}\ }\href@noop {} {\bibfield  {journal} {\bibinfo
  {journal} {Combinatorica}\ }\textbf {\bibinfo {volume} {24}},\ \bibinfo
  {pages} {427--440} (\bibinfo {year} {2004})}\BibitemShut {NoStop}%
\bibitem [{\citenamefont {{The Sage Developers}}(2019)}]{sagemath}%
  \BibitemOpen
  \bibfield  {author} {\bibinfo {author} {\bibnamefont {{The Sage
  Developers}}},\ }\href@noop {} {\emph {\bibinfo {title} {{S}ageMath, the
  {S}age {M}athematics {S}oftware {S}ystem ({V}ersion 8.8)}}} (\bibinfo {year}
  {2019}),\ \bibinfo {note} {{\tt https://www.sagemath.org}}\BibitemShut
  {NoStop}%
\bibitem [{\citenamefont {Lov{\'a}sz}\ and\ \citenamefont
  {Plummer}(2009)}]{lovasz2009matching}%
  \BibitemOpen
  \bibfield  {author} {\bibinfo {author} {\bibfnamefont {L{\'a}szl{\'o}}\
  \bibnamefont {Lov{\'a}sz}}\ and\ \bibinfo {author} {\bibfnamefont
  {Michael~D}\ \bibnamefont {Plummer}},\ }\href@noop {} {\emph {\bibinfo
  {title} {Matching theory}}},\ Vol.\ \bibinfo {volume} {367}\ (\bibinfo
  {publisher} {American Mathematical Society},\ \bibinfo {year}
  {2009})\BibitemShut {NoStop}%
\bibitem [{\citenamefont {Farrell}(1979)}]{farrell1979introduction}%
  \BibitemOpen
  \bibfield  {author} {\bibinfo {author} {\bibfnamefont {Edward~J}\
  \bibnamefont {Farrell}},\ }\bibfield  {title} {\enquote {\bibinfo {title} {An
  introduction to matching polynomials},}\ }\href@noop {} {\bibfield  {journal}
  {\bibinfo  {journal} {Journal of Combinatorial Theory, Series B}\ }\textbf
  {\bibinfo {volume} {27}},\ \bibinfo {pages} {75--86} (\bibinfo {year}
  {1979})}\BibitemShut {NoStop}%
\bibitem [{\citenamefont {Gutman}(1977)}]{gutman1977acyclic}%
  \BibitemOpen
  \bibfield  {author} {\bibinfo {author} {\bibfnamefont {Ivan}\ \bibnamefont
  {Gutman}},\ }\bibfield  {title} {\enquote {\bibinfo {title} {The acyclic
  polynomial of a graph},}\ }\href@noop {} {\bibfield  {journal} {\bibinfo
  {journal} {Publ. Inst. Math.(Beograd)(NS)}\ }\textbf {\bibinfo {volume}
  {22}},\ \bibinfo {pages} {63--69} (\bibinfo {year} {1977})}\BibitemShut
  {NoStop}%
\bibitem [{\citenamefont {Isserlis}(1918)}]{isserlis1918formula}%
  \BibitemOpen
  \bibfield  {author} {\bibinfo {author} {\bibfnamefont {Leon}\ \bibnamefont
  {Isserlis}},\ }\bibfield  {title} {\enquote {\bibinfo {title} {On a formula
  for the product-moment coefficient of any order of a normal frequency
  distribution in any number of variables},}\ }\href@noop {} {\bibfield
  {journal} {\bibinfo  {journal} {Biometrika}\ }\textbf {\bibinfo {volume}
  {12}},\ \bibinfo {pages} {134--139} (\bibinfo {year} {1918})}\BibitemShut
  {NoStop}%
\bibitem [{\citenamefont {Bj{\"o}rklund}\ \emph {et~al.}(2018)\citenamefont
  {Bj{\"o}rklund}, \citenamefont {Gupt},\ and\ \citenamefont
  {Quesada}}]{bjorklund2018faster}%
  \BibitemOpen
  \bibfield  {author} {\bibinfo {author} {\bibfnamefont {Andreas}\ \bibnamefont
  {Bj{\"o}rklund}}, \bibinfo {author} {\bibfnamefont {Brajesh}\ \bibnamefont
  {Gupt}}, \ and\ \bibinfo {author} {\bibfnamefont {Nicol{\'a}s}\ \bibnamefont
  {Quesada}},\ }\bibfield  {title} {\enquote {\bibinfo {title} {A faster
  hafnian formula for complex matrices and its benchmarking on the titan
  supercomputer},}\ }\href@noop {} {\bibfield  {journal} {\bibinfo  {journal}
  {arXiv preprint arXiv:1805.12498}\ } (\bibinfo {year} {2018})}\BibitemShut
  {NoStop}%
\bibitem [{\citenamefont {G{\"a}rtner}\ \emph {et~al.}(2003)\citenamefont
  {G{\"a}rtner}, \citenamefont {Flach},\ and\ \citenamefont
  {Wrobel}}]{gartner2003graph}%
  \BibitemOpen
  \bibfield  {author} {\bibinfo {author} {\bibfnamefont {Thomas}\ \bibnamefont
  {G{\"a}rtner}}, \bibinfo {author} {\bibfnamefont {Peter}\ \bibnamefont
  {Flach}}, \ and\ \bibinfo {author} {\bibfnamefont {Stefan}\ \bibnamefont
  {Wrobel}},\ }\bibfield  {title} {\enquote {\bibinfo {title} {On graph
  kernels: Hardness results and efficient alternatives},}\ }in\ \href@noop {}
  {\emph {\bibinfo {booktitle} {Learning theory and kernel machines}}}\
  (\bibinfo  {publisher} {Springer},\ \bibinfo {year} {2003})\ pp.\ \bibinfo
  {pages} {129--143}\BibitemShut {NoStop}%
\bibitem [{\citenamefont {Kriege}\ and\ \citenamefont
  {Mutzel}(2012)}]{kriege2012subgraph}%
  \BibitemOpen
  \bibfield  {author} {\bibinfo {author} {\bibfnamefont {Nils}\ \bibnamefont
  {Kriege}}\ and\ \bibinfo {author} {\bibfnamefont {Petra}\ \bibnamefont
  {Mutzel}},\ }\bibfield  {title} {\enquote {\bibinfo {title} {Subgraph
  matching kernels for attributed graphs},}\ }\href@noop {} {\bibfield
  {journal} {\bibinfo  {journal} {arXiv preprint arXiv:1206.6483}\ } (\bibinfo
  {year} {2012})}\BibitemShut {NoStop}%
\bibitem [{\citenamefont {Siglidis}\ \emph {et~al.}(2018)\citenamefont
  {Siglidis}, \citenamefont {Nikolentzos}, \citenamefont {Limnios},
  \citenamefont {Giatsidis}, \citenamefont {Skianis},\ and\ \citenamefont
  {Vazirgiannis}}]{siglidis2018grakel}%
  \BibitemOpen
  \bibfield  {author} {\bibinfo {author} {\bibfnamefont {Giannis}\ \bibnamefont
  {Siglidis}}, \bibinfo {author} {\bibfnamefont {Giannis}\ \bibnamefont
  {Nikolentzos}}, \bibinfo {author} {\bibfnamefont {Stratis}\ \bibnamefont
  {Limnios}}, \bibinfo {author} {\bibfnamefont {Christos}\ \bibnamefont
  {Giatsidis}}, \bibinfo {author} {\bibfnamefont {Konstantinos}\ \bibnamefont
  {Skianis}}, \ and\ \bibinfo {author} {\bibfnamefont {Michalis}\ \bibnamefont
  {Vazirgiannis}},\ }\bibfield  {title} {\enquote {\bibinfo {title} {Grakel: A
  graph kernel library in python},}\ }\href@noop {} {\bibfield  {journal}
  {\bibinfo  {journal} {arXiv preprint arXiv:1806.02193}\ } (\bibinfo {year}
  {2018})}\BibitemShut {NoStop}%
\bibitem [{Note5()}]{Note5}%
  \BibitemOpen
  \bibinfo {note} {Experiments were run on IBM's cloud platform using four
  2.8GHz Intel Xeon-IvyBridge Ex (E7-4890-V2-PentadecaCore) processors with 15
  CPU cores each, as well as on Oak Ridge's Titan supercomputer.}\BibitemShut
  {Stop}%
\bibitem [{\citenamefont {Kersting}\ \emph {et~al.}(2016)\citenamefont
  {Kersting}, \citenamefont {Kriege}, \citenamefont {Morris}, \citenamefont
  {Mutzel},\ and\ \citenamefont {Neumann}}]{KKMMN2016}%
  \BibitemOpen
  \bibfield  {author} {\bibinfo {author} {\bibfnamefont {Kristian}\
  \bibnamefont {Kersting}}, \bibinfo {author} {\bibfnamefont {Nils~M.}\
  \bibnamefont {Kriege}}, \bibinfo {author} {\bibfnamefont {Christopher}\
  \bibnamefont {Morris}}, \bibinfo {author} {\bibfnamefont {Petra}\
  \bibnamefont {Mutzel}}, \ and\ \bibinfo {author} {\bibfnamefont {Marion}\
  \bibnamefont {Neumann}},\ }\href {http://graphkernels.cs.tu-dortmund.de}
  {\enquote {\bibinfo {title} {Benchmark data sets for graph kernels},}\ }
  (\bibinfo {year} {2016})\BibitemShut {NoStop}%
\bibitem [{Note6()}]{Note6}%
  \BibitemOpen
  \bibinfo {note} {Standarization of the feature vectors to emphasize their
  mutual differences improved classification accuracy in some cases, but
  deteriorated it in others.}\BibitemShut {Stop}%
\end{thebibliography}
\end{document}